\documentclass[epj]{svjour}
\usepackage{amssymb,amsmath}

\usepackage[usenames]{color}
\usepackage{multirow}
\usepackage{txfonts}
\usepackage[percent]{overpic}
\usepackage{overpic}
\usepackage{bbold}
\usepackage{slashed} 
\usepackage{bm}
\usepackage{booktabs}
\usepackage{graphicx}
\newcommand{\red}[1]{\textcolor{red}{#1}}
\newcommand{\bea}{\begin{eqnarray}}
\newcommand{\eea}{\end{eqnarray}}

\RequirePackage{graphicx}

\begin{document}
\title{Gravitational slingshots around black holes in a binary}
\author{Fan Zhang\inst{1}\inst{2}                     
%
%
\thanks{email: {fnzhang@bnu.edu.cn}}
}
\institute{Gravitational Wave and Cosmology Laboratory, Department of Astronomy, Beijing Normal University, Beijing 100875, China \and Department of Physics and Astronomy, West Virginia University, PO Box 6315, Morgantown, WV 26506, USA}
\date{Received: date / Revised version: date}
%
\abstract{
The speed gain of a test mass from taking a gravitational slingshot around a celestial object (scattering centre) increases with the latter's speed and compactness (stronger deflection of the mass' trajectory becomes possible without it hitting the surface of the object). The black holes (BHs) in a tight binary (consisting of two black holes; we are not considering X-ray binaries), themselves moving at relativistic speeds, represent optimal scattering centres. Therefore, a sub-population of accreting matter particles, swept up into chaotic orbits around a BH binary, might repeatedly take slingshots and become accelerated to ultra-relativistic speeds (as seen by observers on Earth), ultimately escaping from the binary, as well as the fate of being devoured by a BH. The escaped particles can plausibly be observed on Earth as ultra-high-energy cosmic rays (UHECRs). Investigating such a possibility would require general relativistic slingshot formulae due to the high speeds involved and the close encounters with BHs. We derive them in this paper, and show that the percentage gain per slingshot in a particle's Lorentz factor remains undiminished even as the particle energizes up, thus demonstrating that the slingshot mechanism can in principal accelerate particles to extreme energies. 
%
%
}

\maketitle

\section{Introduction}
Ultra-high energy cosmic rays (UHECRs) \cite{1961PhRvL...6..485L,1963PhRvL..10..146L} have garnered tremendous interest since they would suffer less deflection by the galactic and intergalactic magnetic fields, so their paths point roughly back at their sources, potentially helping to identify and elucidate some undoubtedly interesting extremely energetic astrophysical processes, but the nature of which have so far remained an open question. Nevertheless, more recent observations had revealed a wealth of important features in the UHECR spectrum, composition, anisotropy and (lack of) counterparts, that can guide our modelling efforts. For example, the most straightforward top-down approach, where some type of extremely heavy beyond-the-standard-model particle decays into lighter ones whose rest masses add up to only a small fraction of the parent particle's rest mass, is now somewhat disfavoured. Firstly, such processes tend to produce copious amounts of concomitant high energy photons and neutrinos, which are absent observationally \cite{2012APh....35..660K,2017AIPC.1852d0001K}. Also, composition studies show that there should be large amounts of intermediate mass nuclei at the very highest energies \cite{2014PhRvD..90l2005A,2014PhRvD..90l2006A,2015APh....64...49A}, which would be rather perplexing in the decay picture, as one must wonder why ultra-relativistic nucleons (as decay products) all move in the same direction and remain bound inside a nuclei, instead of simply flying off in all directions.

The bottom-up alternative of accelerating initially low energy particles is no less challenging. The macroscopic high energies reaching into hundreds of EeVs means that we need to find an extremely energy-dense environment that can boost the particles to exceptionally high Lorentz factors. 
Thankfully though, the candidate pool for the most powerful accelerators in the universe is limited, and some have already become constrained by observations. For example, while not yet definitive, the popular hypothesis of UHECR being born in active-galactic nuclei (AGN) \cite{2012ApJ...749...63M} is not supported by the lack of correlation between UHECR directional data and the catalogue of powerful AGNs \cite{2010PhRvL.104i1101A,2018ApJ...853L..29A,2018arXiv181101108F}. Furthermore, while $\sim$PeV neutrinos have recently been associated with a Blazar, indicating that AGN jets can accelerate protons and nuclei to ``at least several PeV'' \cite{2018Sci...361.1378I,2018Sci...361..147I}, their relatively low energies in the UHECR context (EeV scale UHECRs will produce neutrinos peaking at $> 100$PeV \cite{2014PhRvD..90b3007M}), when coupled with the stringent X-ray constraints on the peak neutrino flux (the peak is thus forced to be closer to the lower observed energies, and away from the UHECR-relevant values, to be able to satisfy the high observed event rate), in fact implies that the Blazar is ``not a significant UHECR accelerator'' \cite{2018arXiv180704537K} (see also \cite{2018arXiv180704275G} for an independent analysis reaching a similar conclusion). Another leading candidate assuming UHECR being accelerated during gamma-ray bursts \cite{1995PhRvL..75..386W,1995ApJ...449L..37M} is also constrained (for the prompt emission phase) by the lack of accompanying high energy neutrinos \cite{2012Natur.484..351A}. Alternative source sites may thus be worth exploring, and recently, an extremely energetic environment has been observationally proven to exist \cite{2016PhRvL.116f1102A} in abundance \cite{2016PhRvL.116x1103A,2016ApJ...833L...1A,2017PhRvL.118v1101A,2017ApJ...851L..35A,2017PhRvL.119n1101A}, namely stellar mass black holes (BHs) in a tight binary. We thus carry out some initial investigation of the possibility that UHECRs, or a sub-population of them (in which case, some or all of the mechanisms mentioned above may contribute significantly, and the constrains may be due to confusions brought about by the multi-channel complexity), might be born in these systems, out of purely gravitational processes. We caution though, that although we will argue that our present understanding (which is unfortunately next to nothing) of the environment surrounding binary BHs (BBHs) does not exclude our proposed mechanism as a significant contributor to UHECRs (i.e., we will show that optimistic parameter estimates exist, but we don't have sufficient evidence to claim that they are the most probable), the analysis presented in this paper is severely limited by technical difficulties, and we in no way claim that it definitively would play such a role. Instead, we wish only to solicit interest for an acceleration mechanism that we think deserves closer scrutiny.  

We begin by motivating gravitational slingshots or gravity assist as an acceleration mechanism through more qualitative considerations in Sec.~\ref{sec:Qualitative}, before deriving quantitative general relativistic slingshot formulae in Sec.~\ref{sec:Ana}, which shows that there is no degradation in the efficiency of the slingshots even as the cosmic ray (CR) particles already possess very high energies, making acceleration to the exceptional energies of the UHECRs possible. We note that numerical studies tracking pressureless matter flow around a binary have been carried out by Ref.~\cite{2010ApJ...711L..89V}, in order to study electromagnetic counterparts to gravitational waves (GWs). That study already demonstrated that even the bulk of that matter could be accelerated to high Lorentz factors, but here we are more interested in the extremely energetic outliers, so an analytical study, beyond providing more insight, is perhaps more convenient, since numerical studies would need to track an extremely large number of particles to achieve a decent sample size of the outliers. In regard to other UHECR phenomenology, due to the relatively low event rate, directional deflection by magnetic fields and the uncertainty about particle physics at such high energies, definitive observational inferences are relatively few. Nevertheless, several broad stroke features appear to have emerged, and we will discuss the compatibility of the proposed repeated slingshots (RS) mechanism with them. These discussions will be blended into Secs.~\ref{sec:Qualitative} and \ref{sec:Ana}. We then conclude and motivate further studies in Sec.~\ref{sec:Conclusion}. In this paper, we work in geometrized units with $G=c=\epsilon_0=1$. Within that unit, we use solar mass $M_{\odot}$ as the fundamental length unit, which equates to $1.5$km (or $5\times 10^{-6}$s) when translated back to cgs or SI units. The metric convention we adopt is $(-+++)$.

\section{Gravitational acceleration} \label{sec:Qualitative}
\subsection{Motivating gravity-based mechanisms} \label{sec:SlingsIntro}
One consideration motivating us to look at gravity-based acceleration mechanisms is the fact that the alternatives suffer from the Abraham-Lorentz-Dirac damping (see e.g., \cite{1938RSPSA.167..148D,1999gr.qc....12045P}). 
This is an under-scrutinized separate and additional radiation loss channel to the ones, such as synchrotron radiation, that's usually examined in the UHECR production context (see e.g., \cite{1995PhRvL..75..386W}), and is more difficult to circumvent because it is an intrinsic by-product of the energization process itself. Namely, the act of accelerating charged particles would in itself induce electromagnetic radiations (whose power is proportional to the acceleration squared, see e.g., Eq.~25 in \cite{1997PhRvD..56.3381Q}), and so the particle experiences a decelerating back-reaction. This means while energy is being pumped into the charged particles, e.g., by some electrostatic field \footnote{Note that magnetic field does no work, so ultimately electric fields have to be involved, but the to-be-accelerated ions themselves or electron-positron pair production tends to short them out, so long-lived electrostatic fields may not be easy to sustain.}, or through the electromagnetic scatterings in the Fermi processes\footnote{In this case, the Abraham-Lorentz-Dirac damping would manifest as extra photons coming off when the charged particles quickly turn around when hitting a magnetic scattering centre, which must partake in the conservation of energy and momentum computations. This complication is usually not taken into account, but should be important in the UHECR context, since the scenario of ultra-relativistic particles taking a near U-turn without emitting any photon seems unlikely.} \cite{1949PhRv...75.1169F,1977ICRC...11..132A,1978MNRAS.182..147B,1978ApJ...221L..29B}, 
some of it inevitably leaks out, and the more intense the injection, the greater proportion of the energy is lost. Therefore, if one wants to rapidly boost the UHECR nuclei's speed to ultra-relativistic values, they must be prepared to accept a diminished efficiency, enhancing the demands on the force field intensity at the source site. On the other hand, as compared to electromagnetism (EM), gravity offers an interesting loss-free alternative. Because the nuclei in a vacuum gravitational field would simply follow geodesics and are freely falling (at leading order), their radiation reaction is suppressed (see \cite{1997PhRvD..56.3381Q} for precise formulae where higher order corrections are also included). An intuitive portrait of this effect must be somewhat nuanced, since the statement of whether a charged particle radiates is observer dependent \cite{1955RSPSA.229..416B,1963AnPhy..22..169R}. Specifically, a freely falling charge does not radiate as seen by a comoving inertial observer, but it radiates when observed by a stationary/accelerating observer \cite{Gron:2012pr}. Nevertheless, as is clear in the simpler case of an accelerating observer seeing radiation coming off a static charge in a flat spacetime, such radiation does not necessarily imply a draining of energy from the particle. We can construct a more gauge invariant description as follows: at any time during the free-falling motion of a charged particle $A$, we can place a free-falling charge-neutral particle $B$ at the same location, with a matching instantaneous velocity, so that $B$ corresponds to the aforementioned comoving inertial observer. Since this observer will not see particle $A$ radiating (thus suffers no radiation reaction), it will not find any acceleration of $A$ relative to herself, and so $A$ and $B$ will remain in lock-step to the next moment in time, where the same analysis can again be applied. Therefore, the two particles will share the same geodetic trajectory, with $A$ suffering no extra loss of energy due to it carrying an electric charge.

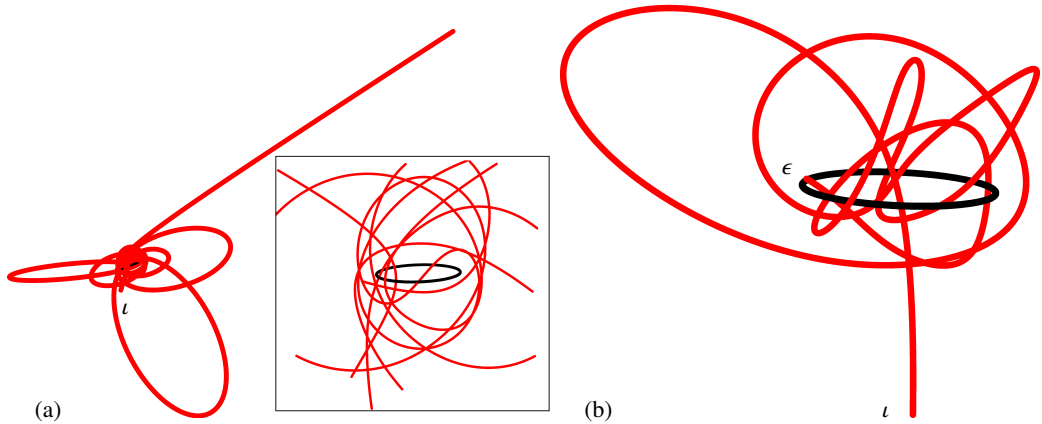
\begin{figure}
  \centering
\begin{overpic}[width=0.4\columnwidth]  {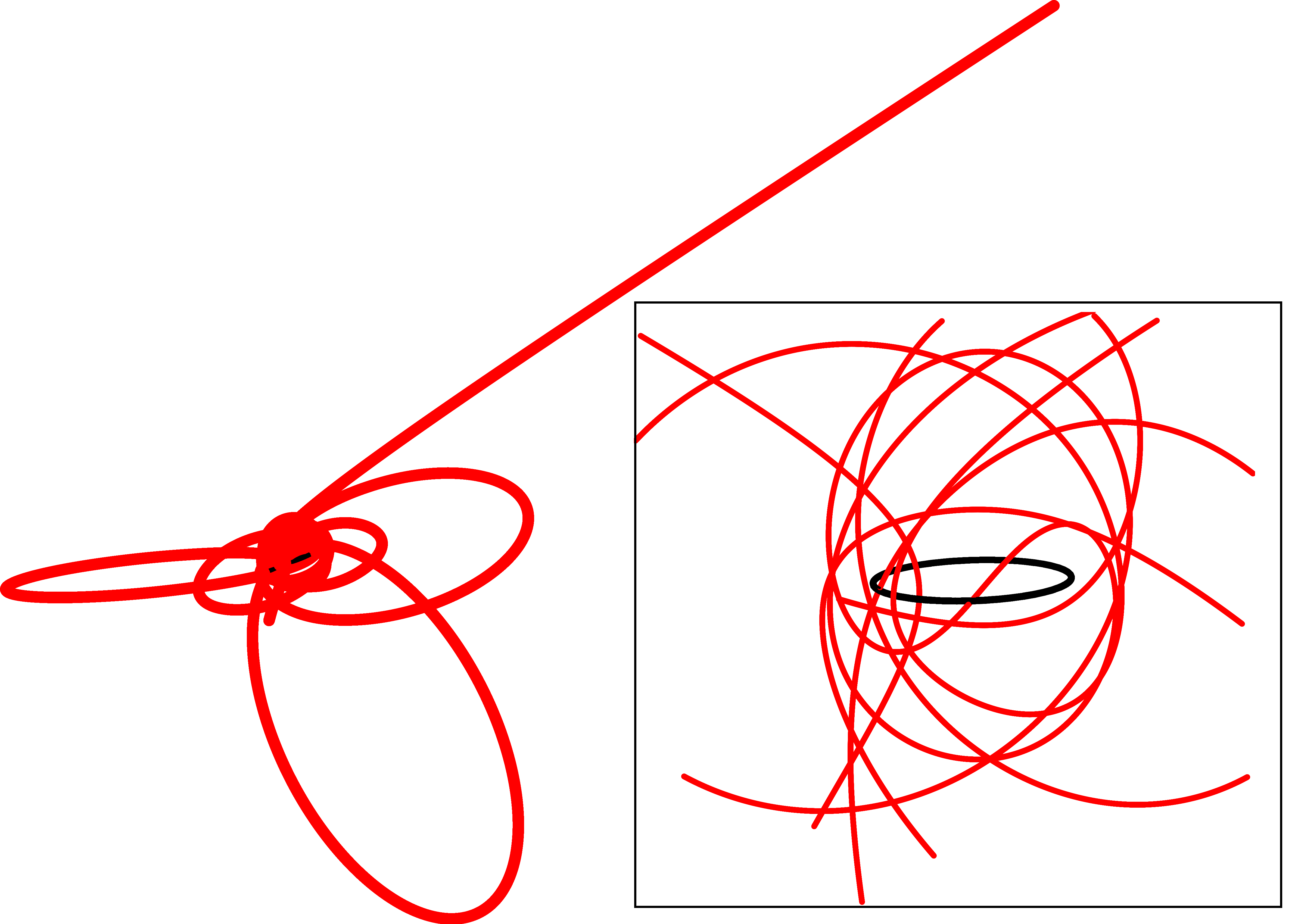}
\put(21,19){$\iota$}
\put(5,0){(a)}
\end{overpic}
\begin{overpic}[width=0.35\columnwidth]  {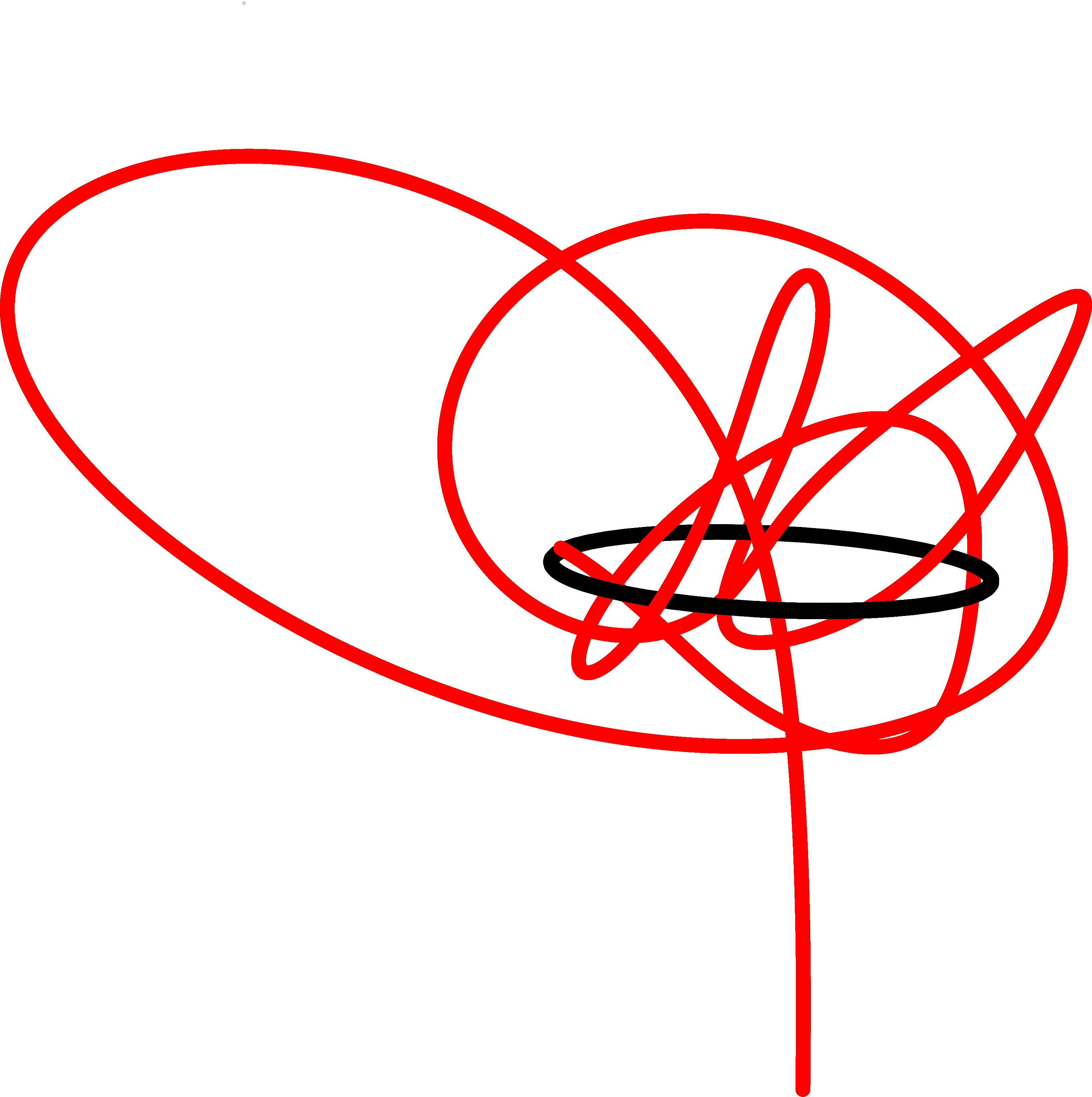}
\put(67,0){$\iota$}
\put(46,51){$\epsilon$}
\put(5,0){(b)}
\end{overpic}
  \caption{A couple of particle trajectories (red) integrated around a binary of equal-mass BHs, using Newtonian gravity. The black circle indicates the orbits of the BHs, and its intersection with the trajectories does not always mean that the particle falls into a BH, as the crossing can be temporally mismatched. The locations of the starting positions are marked out with $\iota$. (a): A particle that takes a number of slingshots, become accelerated to high speeds and escape the binary. The inset is a zoom-in on the central region (not the same viewing angle as the main figure). Note this geodesic is an arbitrary generic one, from a sample of a few integrations that we performed for illustration purpose only. It is not specially sought to produce the large number of slingshots needed to produce UHECRs, which if plotted would in any case cover the figure too densely and not be visually helpful. (b): Another particle that experiences many slingshots but not always in an accelerating configuration, and end up falling into a BH at $\epsilon$. }
	\label{fig:Trajs}
\end{figure}
 
The celestial objects possessing the most intense gravitational fields are the BHs, but which type to concentrate on? There are both X-ray binaries and a supermassive BH in our own galaxy, at distances of at most a few kpc from us. So galactic sources would likely dominate the UHECR if single BHs are the sources. This is the opposite to what is seen \cite{2010PhRvL.104i1101A,2017Sci...357.1266.}. Looking beyond single BHs, binaries are the obvious next category. The occurrence of supermassive BH binaries in the nearby universe is too low to be interesting (pulsar-timing arrays set strong limits \cite{2013Sci...342..334S,2016ApJ...821...13A,2015MNRAS.453.2576L}), but it is observationally proven that there exists stellar mass BBHs in the local universe (but no tight advanced-stage pairs in the Milky Way, otherwise LIGO and VIRGO would have seen them even when they are still in the late inspiral phase, so stellar BBH sources are consistent with no local dominance in the UHECRs). 
In addition, there is now tentative evidence of UHECR directional correlation with starburst galaxies \cite{2018arXiv180106160T}, where BBHs are likely more abundant than average (see e.g., discussions in \cite{2017ApJ...836...50L}).
We will thus examine the stellar mass BBH scenario in more details, since it also possesses the required energy density, and is dynamic enough to allow for interesting kinetics to inject energy into accreting matter particles.

Specifically, it provides the setting for an extreme version of the gravity assist process commonly employed in deep space explorations. The particles taking a slingshot off of the extremely fast moving BHs in a tight binary will receive a tug that relativistically boost its speed. Furthermore, when a particle is travelling in-between the BHs, the gravitational pull due to the holes point in roughly opposite directions, so the particle is not necessarily inclined to quickly fall into either one, especially if the holes are spinning rapidly (infalling particles are deflected to help enforce cosmic censorship \cite{1974AnPhy..82..548W}). This means that a subset of particles could negotiate the complex tomography of the gravitational potential when traversing the interior regions of the binary, without quickly falling into either hole (see Fig.~\ref{fig:Trajs} for pictorial examples of quite generic chaotic orbits followed by particles making their ways around a binary). Opportunities exist then, for a further subset of fortunate particles to take multiple slingshots and be accelerated to ultra-relativistic speeds (become more light-like), making them even more difficult for the gravity of either hole to grab onto (i.e., with more slingshots, the particles become less likely to fall into the BHs, and more likely to participate in further slingshots), eventually allowing them to escape the binary region all together. A corollary of this discussion is that the RS mechanism does not operate effectively on massless particles like photons, that are already moving at the speed of light, and will simply escape quickly. So there won't be any accompanying ultra-high energy photons to the UHECRs, in agreement with observations \cite{2012APh....35..660K,2017AIPC.1852d0001K,2017JCAP...04..009A}. 

\subsection{Composition} \label{sec:composition}
Being gravity-based, the RS process operates independently of the rest mass of the particle being accelerated (so long as that mass is not zero), so it will accelerate different nuclei to the same Lorentz factor, resulting in nuclei with mass number $A$ having $A$ times the energy of a proton. Such a migration into heavier nuclei at higher energies is supported by observations \cite{2010PhRvL.104i1101A,2014PhRvD..90l2005A,2014PhRvD..90l2006A,2015APh....64...49A,2016uhec.confa0016A,2017AIPC.1852d0001K}, and is more pronounced with RS than EM alternatives, which are often assumed to be rigidity-dependent (e.g., through considerations of magnetic confinement to the acceleration region), with total energy proportional to the atomic number $Z$ instead of $A$ \cite{2011APh....34..620A}. 

This provides us with a way to distinguish between gravitational and EM acceleration mechanisms, and hints that scaling according to $A$ is more appropriate have already begun to emerge from data. The most direct and objective method is to examine the composition trend against energy, as extracted without referencing any source models (but dependent on hadronic interaction models). With either mechanisms, the flux density spectra of different elements will simply share the same spectral shape, but
are shifted against each other along the logarithmic energy axis by either $\log Z$ or $\log{A}$, and weighted by their relative abundances at source. One can then reasonably expect that the percentage composition of an element will peak near the top of its own flux distribution, provided that the drop-off in the high energy end of the shared spectral shape is fairly sharp (peak-to-cutoff distance is small as compared to $\log Z$ or $\log A$), because otherwise the tails of the more abundance lighter elements will protrude into the peak regions of heavier elements, still dominating in fractional presence due to their shear overwhelming total population, thus pushing peak fraction for the rarer heavier elements further into higher energies. Fortunately, the small observed variance at the highest energies in 
the depth of atmospheric shower max $X_{\rm max}$ restricts the simultaneous presence of elements of very different $A$ at the same energy (and especially the presence of protons that tend to cause the most variability -- the variations partially average out for heavier elements as more incident nucleons are involved), so the tail cannot be too fat. Furthermore, the abundances of hydrogen and helium would be relatively close (an order of magnitude difference in our solar neighbourhood, but nitrogen is four orders of magnitudes rarer than hydrogen, see \cite{2003ApJ...591.1220L}), so the required sharpness of cutoff is much relaxed if we only compare the peak fraction locations of these two lightest elements.  

\begin{figure}
  \centering
\begin{overpic}[width=0.49\columnwidth]  {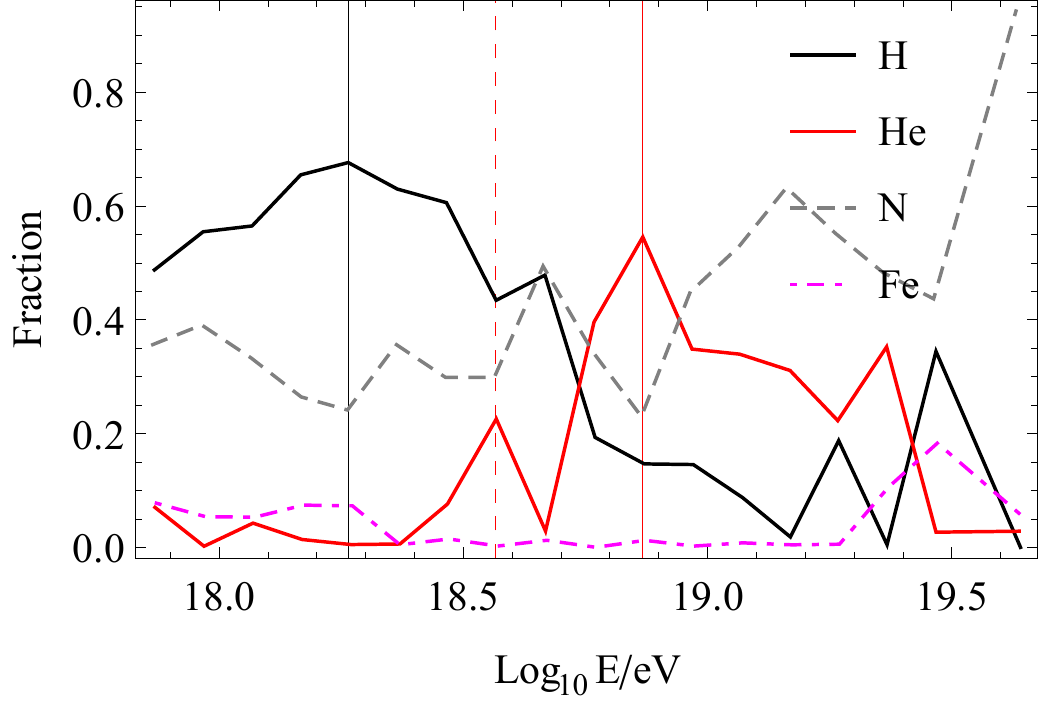}
\end{overpic}
  \caption{The fractional composition of hydrogen, helium, nitrogen (representing the C/N/O intermediate mass group) and iron in the UHECR, assembled with data taken from Fig.~4 of \cite{2014PhRvD..90l2006A}. The energy corresponding to peak fraction of hydrogen is indicated with a solid black vertical line, and those correspondingly predicted for helium are shown as solid (shifted from the black line by $\log_{10}A$ for gravitational acceleration) and dashed (shifted by $\log_{10}Z$ for EM acceleration) red vertical lines.}
	\label{fig:Comp}
\end{figure}

We carry out this investigation by collecting in Fig.~\ref{fig:Comp} the composition fittings to $X_{\rm max}$ produced by \cite{2014PhRvD..90l2006A} (we only consider the EPOS-LHC \cite{2006PhRvC..74d4902W} hadronic interaction model that incorporates the $7$TeV LHC data and provides the best fit to observational data, see the conclusion section of \cite{2014PhRvD..90l2006A}), and the result appears to favour the $A$-dependent gravitational acceleration mechanisms. 
Furthermore, it is worth noting that their fits exclude a proton plus iron composition, and in fact excludes significant iron contributions even when acceptable fits are produced by introducing intermediate elements (see the magenta dot-dashed line in Fig.~\ref{fig:Comp}). This is somewhat strange as iron nuclei are very stable against photodisintegration \cite{2007APh....27...61A} and are fairly abundant in the universe (similar abundance to nitrogen in the solar neighbourhood \cite{2003ApJ...591.1220L}). It is however reasonable in the context of the RS mechanism, since the iron cores of the progenitor stars should have all collapsed into BHs, while the debris forming an accretion disk that injects UHECR candidate particles into the binary region would instead originate primarily from the outer layers, consisting of lighter elements, that have been blow out during the hypernovae. 

We further note that the heaviest and most energetic UHECR constituents being in the C/N/O group has a rather desirable consequence, since then the most important attenuating factor at the highest energies becomes photodisintegration \cite{1976ApJ...205..638P}, whose associated attenuation length exhibits a sharp turn at a Lorentz factor of $3\times10^9$, beyond which it begins to trend more steeply lower against energy (see the left panel of Fig.~1 in \cite{2015PhRvD..92f3014H}). This threshold Lorentz factor equates to an energy of around $40$EeV for the C/N/O group. Because the number of sources scales as distance cubed (for a uniform distribution), the slope of the resulting dropoff in the flux density would be further amplified as compared to this one in the distance horizon, so this effect could plausibly explain the sharp cutoff in the UHECR flux density spectrum, occurring also at $40$EeV. In other words, the cutoff may still be a propagation effect, just not due to the Greisen-Zatsepin-Kuzmin (GZK) limit \cite{1966PhRvL..16..748G,1966JETPL...4...78Z}, which only has a chance to exert itself with protons (energy limits for the other elements scale as $A$, and goes above $40$EeV), but there are probably hardly any protons at the highest UHECR energies (see Fig.~\ref{fig:Comp}). This scenario would be consistent with the lack of observed ultra-high energy neutral GZK by-products \cite{2017AIPC.1852d0001K}. 

Furthermore, photodisintegration for the heavier elements (see the curve segments corresponding to UHECR energies in Fig.~1 of \cite{2015PhRvD..92f3014H}) and pair production for protons \cite{2018PrPNP..98...85M} both set the maximum distance to the sources of the highest energy UHECRs at around $1$Gpc (i.e., the sources should be within the local universe), but this does not prevent CRs originating further out from reaching us altogether. Photodisintegration does not reduce the Lorentz factors of the composite nuclei that it breaks up, so the CRs from cosmological sources ($\gtrsim 1$Gpc) will all turn into ultra-relativistic protons, which will also become partially energy-depleted through the pair production effect until their energies fall to around $5$EeV, below which the pair production attenuation length rises to tens of Gpc (see Fig.~8 of \cite{2018PrPNP..98...85M}; and also discussions therein along similar lines, although it is the enhanced photodisintegration at EM-active sources that was envisioned there). Consequently, cosmological sources could possibly provide the (extra-galactic) pure proton injection necessary to explain the ankle feature in the UHECR spectrum (at around $5$EeV), as required by the dip model \cite{1985PhRvD..31..564H,1988A&A...199....1B}. 

In summary, the RS mechanism could plausibly fit into a self-consistent picture, for which the mass composition versus energy relation, to be revealed in much greater clarity by the upcoming AugerPrime \cite{2017AIPC.1852d0001K} and TAx4 \cite{2016NPPP..279..145S} upgrades, will provide a more definitive test. Even at the present time though, the tentative hints are perhaps already sufficiently interesting to warrant a preliminary investigation of the basic properties of the RS mechanism, which we turn to in the next section.

\section{Preliminary quantitative analysis}\label{sec:Ana}
\subsection{Newtonian slingshots} \label{sec:Newtonian}
Although not quite sufficient in accuracy, some salient features of the RS process can be gleaned from a back-of-the-envelop Newtonian calculation, which gives that during each slingshot encounter, a particle with an initial speed $v_i$, approaching a BH of speed $v_{\rm BH}$, with a deviation angle $\varphi$ from the head-on direction, will emerge with speed 
\bea \label{eq:NewtonSling}
v_f = \left(v_i+2v_{\rm BH}\right)\sqrt{1-\frac{4 v_{\rm BH} v_i(1-\cos\varphi)}{(v_i+2v_{\rm BH})^2}}\,,
\eea 
provided it emerges in a symmetric fashion also with angle $\varphi$ to the BH's velocity.
The maximum gain is achieved when the particle reverses course ($\varphi=0$), in which case it picks up twice the speed of the BH, much like a small billiard ball hitting a much larger one head-on, bouncing back elastically and getting an optimal boost to its speed at twice the speed of the larger ball. With slingshots however, the particle cannot hit the scattering centre face-on and will instead swing around the back of it (see Fig.~\ref{fig:GeoEndsPics} below), so the head-on configuration is only possible if the particles can approach sufficiently close to the gravitating centre to experience enough gravity to severely bend (fold) its trajectory. For example, with a spacecraft doing slingshots around a planet, the limit is how deep it can skim the edge of the atmosphere without crashing into it (Eq.~\eqref{eq:NewtonSling} does not take such issues into account, and should really be seen as making the approximation where the scattering centre is a point mass). 
In other words, the key for achieving high speeds through slingshots is that the scattering centres should be fast-moving compact objects, and BHs in a tight binary satisfy this criteria optimally. 

\begin{figure}
  \centering
\begin{overpic}[width=0.79\columnwidth]  {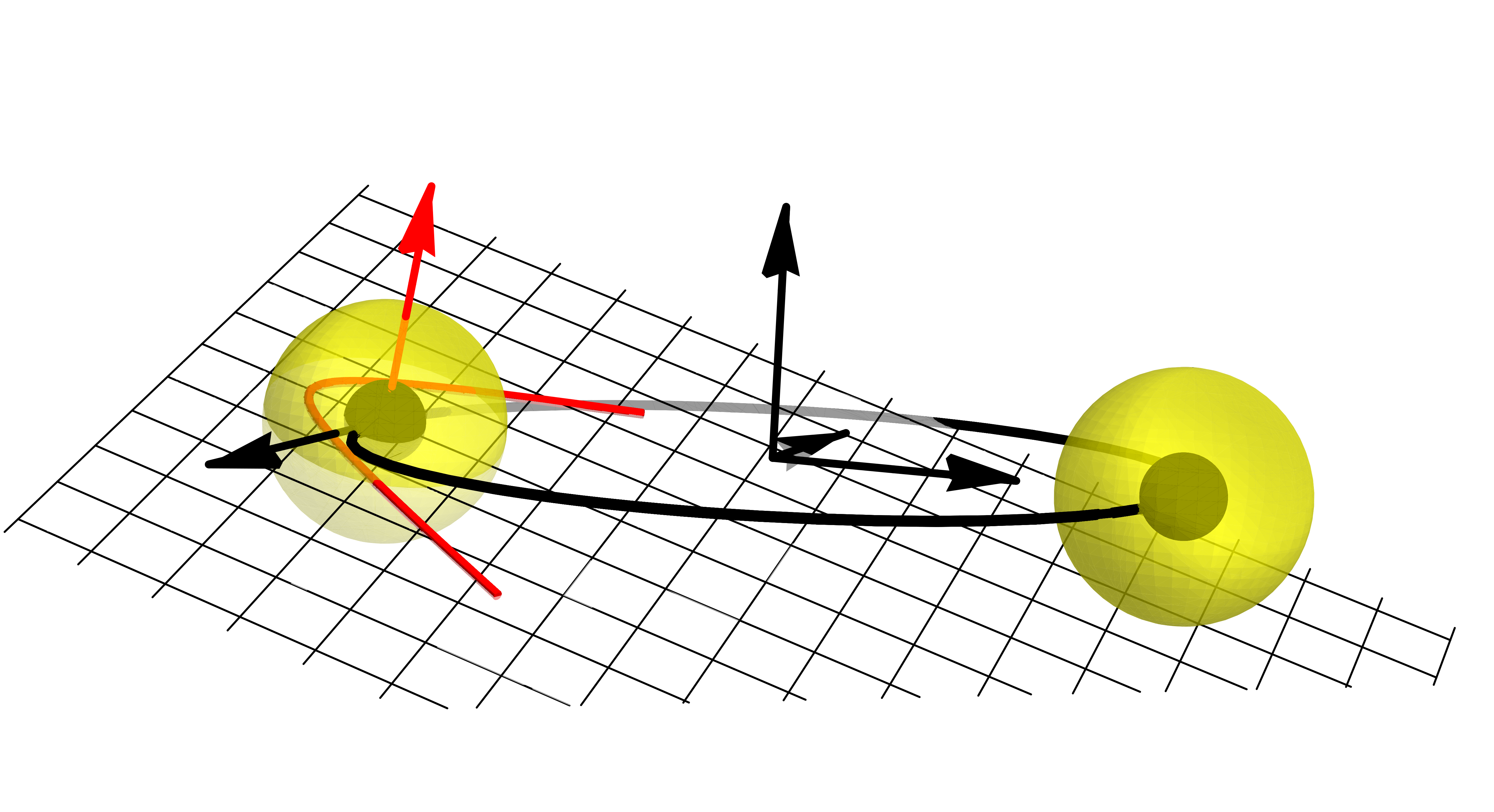}
\put(60,35){NZ}
\put(77,25){IZ}
\put(69,21){x}
\put(57,25){y}
\put(52,41){z}
\put(28,43){\red{$\tilde{z}$}}
\put(11,21){${\bf v}_A$}
\end{overpic}
  \caption{Schematic demonstration of the demarcation of zones within a binary environment, and the various quantities involved in a single slingshot. The black spheres are the event horizons of two equal-mass BHs. The yellow spheres, located at the ISCO of each individual BH, are the boundaries separating the Inner Zones (IZs) and the Near Zone (NZ). At much further out from the binary (not shown in the figure), there is also a Wave Zone, in which the more slowly declining gravitational wave component of the curvature tensor becomes dominant over the Coulomb component, and time retardation effects become important. The black arrows at the centre are the binary frame Cartesian basis vectors. The red arrow is the $\tilde{z}$ axis of the comoving frame adapted to the geodesic (red curve), whose plane of residence (meshed grid) is orthogonal to the red arrow. The velocity ${\bf v}_A$ of the BH in the binary frame is also shown. 
  }
	\label{fig:Zones}
\end{figure}
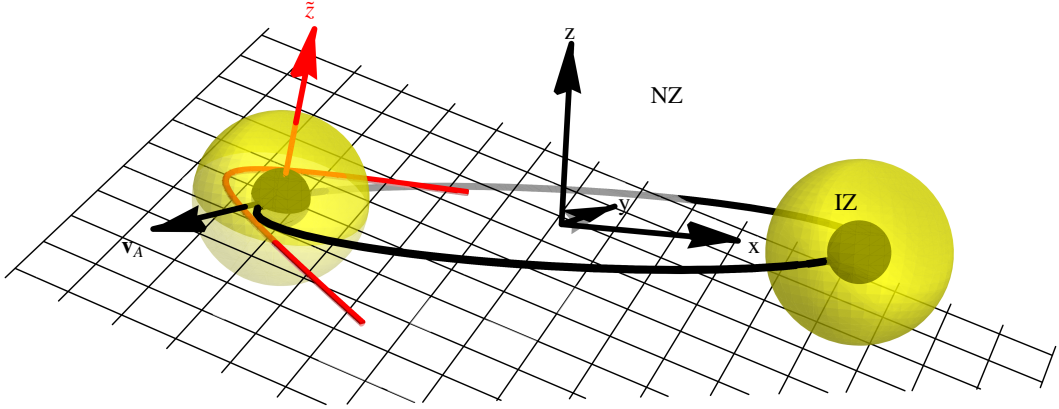

Although useful for intuition building, the simple Newtonian expression \eqref{eq:NewtonSling} severely over-estimates the efficiency of the slingshots. Take for example the masses of the two BHs to be $M_A=1/2M_{\rm tot}$ (where $A=1,2$ labels the two holes), and their separation at $b=2000 M_{\rm tot}$ (still quite far from merger), the Keplerian orbital angular velocity 
\begin{align} \label{eq:Kepler}
\Omega = \frac{1}{b}\left(\frac{M_{\rm tot}}{b}\right)^{1/2}\,,
\end{align}
gives the speed of each BH at $v_{\rm BH}\approx 0.01$c (we will occasionally keep $c$ around for emphasis, even though it has a value of unity in our units). This means that the Newtonian formula predicts that a mere $50$ slingshot episodes at optimal configuration will launch particles to superluminal speeds (and this number drops to five with $b=20M_{\rm tot}$, which corresponds to the late inspiral phase but the BHs are still well outside of each other's innermost stable circular orbits or ISCO). A special relativistic extension to the elastic scattering problem, capping the maximum speed, would also not suffice.
Since the most effective near-head-on slingshots would involve particles passing very close to the BHs, near or inside the Inner Zones (see Fig.~\ref{fig:Zones} and Fig.~\ref{fig:GeoEndsPics}), where general relativistic effects are important. For example, the general relativistic correction to the effective radial potential (see the last term in Eq.~\eqref{eq:GRPot} below, and Fig.~\ref{fig:Potential}) causes some geodesics with small impact parameters to fall into the BH. Or in other words, the collisions would not always be elastic, a situation that is similar but dynamically more complicated than the planetary atmosphere example discussed in the last paragraph. To obtain more appropriate estimates then, we will have to derive fully general relativistic expressions that takes the details of actual orbits (i.e., the viability of actually achieving any prescribed $\varphi$) into account. 

Perhaps due to their lack of practical utility in the foreseeable future however, general relativistic treatments of slingshot geodesics, as far as the author could ascertain, have not become an extensively investigated subject. While low order post-Newtonian expressions (on angle deflection, not speed gain) valid near our sun is available \cite{2001PhRvL..86.2942L}, a fully exact general relativistic treatment needed for slingshots close to the event horizon of a BH is currently missing. We fill this gap in the next section, taking advantage of the fact that for Schwarzschild BHs, there exists analytical solutions to the geodesics in the comoving frame, and a simple boost of the coordinates allows us to translate the results to a resting frame.

\subsection{Relativistic slingshots}\label{sec:SingleSling}
\begin{figure}
  \centering
\begin{overpic}[width=0.24\columnwidth]  {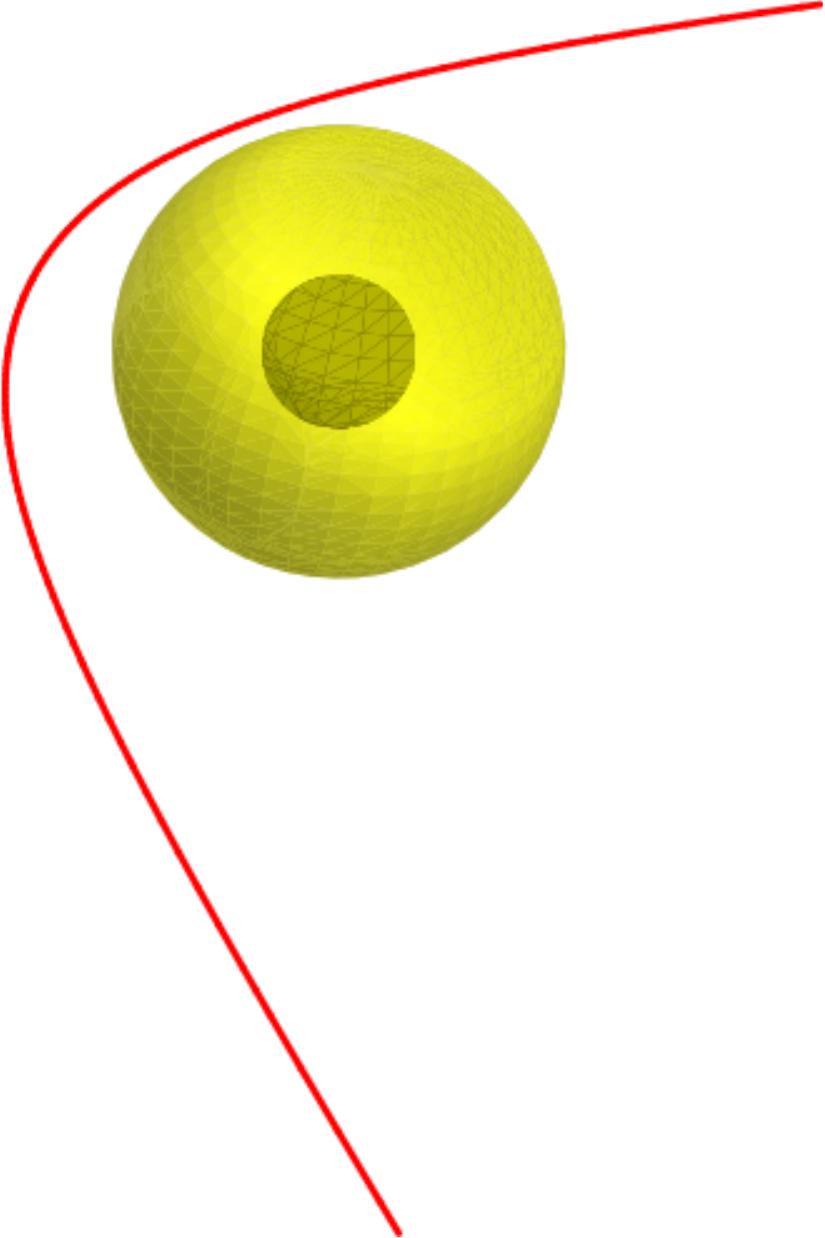}
\put(5,5){(a)}
\end{overpic}
\begin{overpic}[width=0.34\columnwidth]  {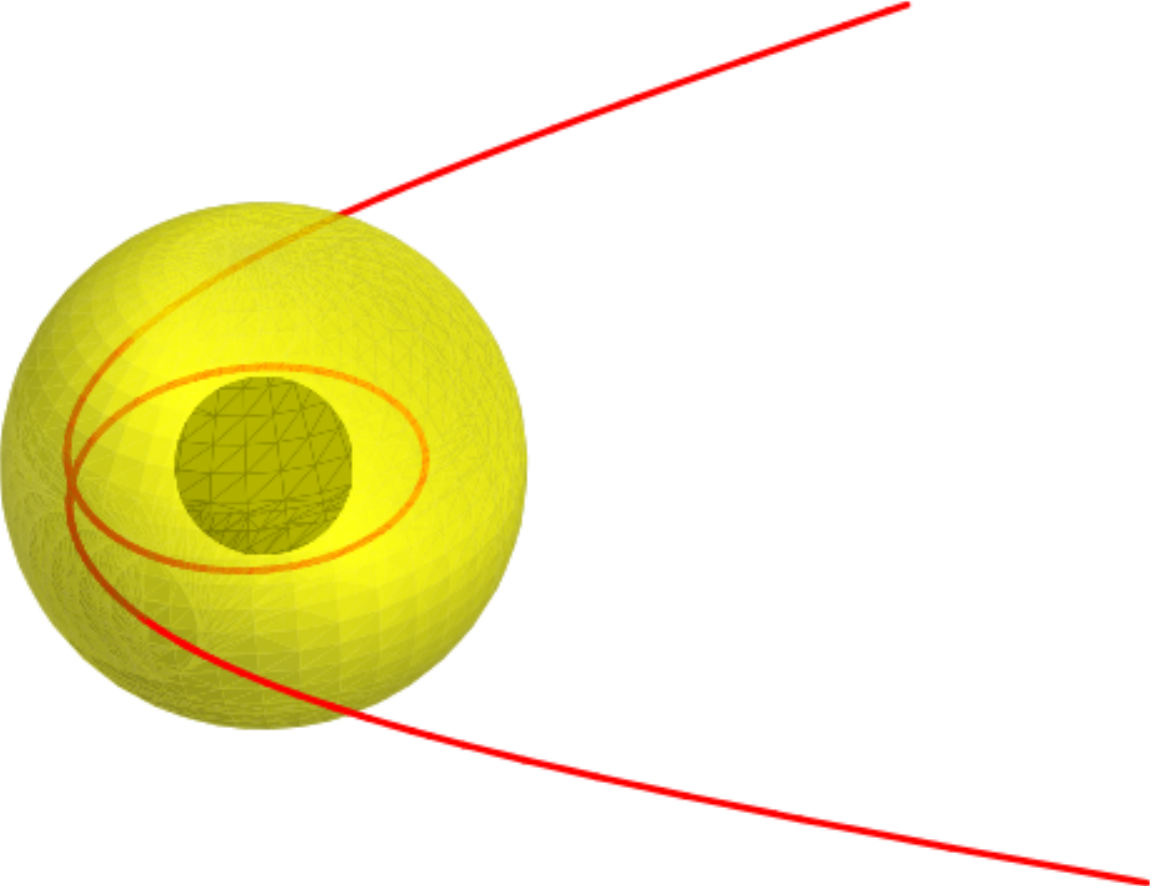}
\put(5,5){(b)}
\end{overpic}
  \caption{The geodesics with (a): $\mathcal{L}$ much larger than $\mathcal{L}^{\rm min}$ and (b):$\mathcal{L}$ slightly larger than $\mathcal{L}^{\rm min}$, respectively. The black spheres once again represent the event horizons, and the yellow surfaces the boundaries of the IZs. The trajectories are plotted as seen in the comoving frame, so the locations of the holes are fixed.}
	\label{fig:GeoEndsPics}
\end{figure}

To begin with, it is useful to quickly re-examine why gravity assist works in the relativistic context, even as the particles are not really accelerated in the traditional sense, but simply follow geodesics. Recall that the geodesic equation ($q^{a}$ denotes the four-coordinates of the geodesic, and $\tau$ its proper time)
\bea\label{eq:GeoGen}
\frac{d^2 q^a}{d\tau^2} +\Gamma^{a}_{bc} \frac{d q^b}{d\tau}\frac{d q^c}{d\tau} = 0
\eea
preserves the norm of the tangent vector to the geodesic (as it is parallelly transported)
\bea
n_{\tau} \equiv \sqrt{-g_{ab}\frac{dq^a}{d\tau}\frac{dq^b}{d\tau}}\,,
\eea
but not the individual component $dt/d\tau$ (it is mathematically impossible to conserve this time dilation factor without imposing additional gauge conditions, since then one is always at liberty to redefine $t$, but not $\tau$), so the amplitude of the coordinate velocity (as measured by cosmic ray observers on Earth for example)
\bea
n_{t} \equiv \sqrt{-g_{ab}\frac{dq^a}{dt}\frac{dq^b}{dt}}=\sqrt{-g_{ab}\frac{dq^a}{d\tau}\frac{dq^b}{d\tau} \left(\frac{d\tau}{dt}\right)^2}=\frac{d\tau}{dt} n_{\tau}
\eea
is not conserved along geodesics, making acceleration (as seen by observers on Earth) by gravitational slingshots possible. In other words, one may see the acceleration as due to a residual time dilation picked up during a close encounter with a BH, which is obviously difficult to achieve in the comoving frame where the scattering centre is a static BH, because there would be a symmetry between the incoming and outgoing legs of a particle's trajectory. However, such symmetries are broken when the BH is moving (or spinning). In other words, the frame of reference in which we measure the energies matters, a fact that is consistent with Newtonian slingshot intuitions ($v_f=v_i$ when $v_{\rm BH}=0$ in Eq.~\eqref{eq:NewtonSling}), and incidentally also somewhat similar to the Fermi process, where there is no electric field thus no acceleration in the comoving frame of the magnetic scattering centres.

For our study, the motion of the BH is due to it being in a binary, whose metric does not admit an exact analytical solution. Fortunately, the most important segment of a slingshot trajectory is that part in the close vicinity of individual holes. To traverse the IZ of a $30M_{\odot}$ BH, 
a particle with a low speed of $0.1$c would only take about $0.02$s, while the BHs will circle each other on $0.2$s timescales even during the very late inspiral phase when $b=20M_{\rm tot}$ ($3$min when $b=2000M_{\rm tot}$), so one can argue that, as far as the slingshot process is concerned, it is reasonable to approximate each hole as a boosted Schwarzschild solution in uniform straight-line motion, with a velocity matching its instantaneous velocity in the binary (i.e., the orbital motion being in actuality circular is not relevant during each slingshot episode). One of course should be concerned that the $0.02$s is a Newtonian estimate, and gravitational redshift may prolong it, depending on how close the trajectory gets to the BH. Using the lapse function of the Schwarzschild BH however, we see that the two timescales become comparable only when the particle gets within $0.02M_{A}$ ($2.5\times 10^{-8}M_A$ for $b=2000M_{\rm tot}$) of the event horizon, and spends the entirety of around $0.02$s of its proper time there. Such extreme situations are not necessary for our discussion below. We are then in a position where analytical expressions can be produced, by first considering geodesics in the comoving frame of the BH, and then translating the results into a ``resting'' frame where the BH is moving with the aforementioned linear velocity (see Sec.~\ref{sec:Assist} for further details).

\subsubsection{The geodesics}
The analytic expressions of the geodesics in the comoving frame of a Schwarzschild BH are well known (see \cite{Whittaker,Hagihara}), but in the interest of a more self-contained discussion, we briefly summarize the main steps of their derivation. We also derive some explicit expressions that are not usually listed in standard references, which tend to concentrate more on circular or precessing elliptic-like orbits rather than the hyperbolic-like ones that are relevant for us. 

\begin{figure}
  \centering
\begin{overpic}[width=0.40\columnwidth]  {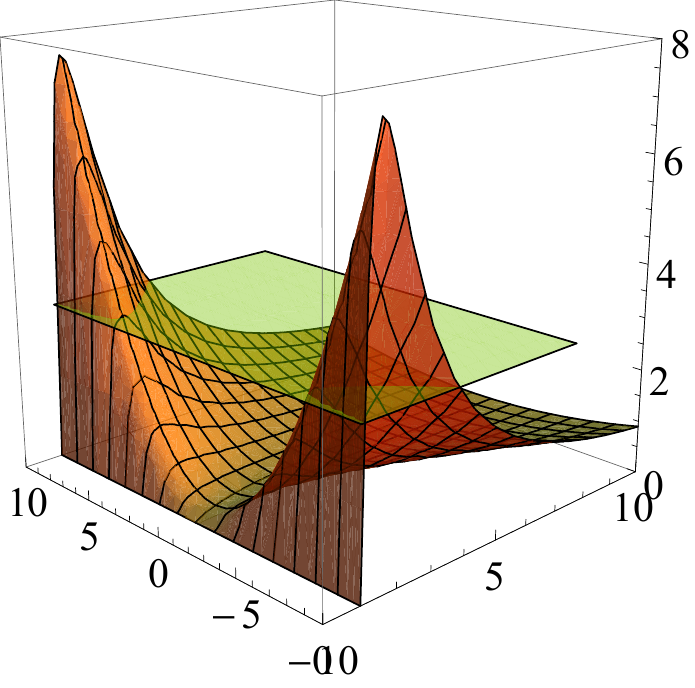}
\put(0,10){$\mathcal{L}/M_{\rm tot}$}
\put(80,10){$\tilde{r}_A/M_{\rm tot}$}
\put(100,60){$V(\tilde{r}_A)$}
\end{overpic}
  \caption{The effective radial potential $V/2$ is plotted as a function of $\tilde{r}_A$ and $\mathcal{L}$, as the meshed surface. An arbitrary demonstrative total radial motion energy $\mathcal{E}^2/2$ is shown as a green transparent smooth surface. A particle whose $\mathcal{E}^2/2$ surface slices into the $V/2$ surface at the $\mathcal{L}$ value that it possesses will run into a potential barrier when coming towards the BH, and move back out without falling into the hole. These are the particles relevant for our slingshot consideration. 
  }
	\label{fig:Potential}
\end{figure}

Let $(\tilde{t}_A,\tilde{r}_A,\tilde{\theta}_A,\tilde{\phi}_A)$ denote the comoving Schwarzschild coordinates associated with BH $A$, then along any timelike geodesic parameterized by proper time $\tau$, we have 
\begin{align} \label{eq:SchMetric}
ds^2 &= -d\tau^2 \notag \\ 
&=- f(\tilde{r}_A)d\tilde{t}_A^2 + \frac{1}{f(\tilde{r}_A)}d\tilde{r}_A^2 +\tilde{r}_A^2\left( d\tilde{\theta}_A^2 + \sin^2\tilde{\theta}_A d\tilde{\phi}_A^2\right)\,,
\end{align}
where 
\bea
f(\tilde{r}_A) = 1-\frac{2M_A}{\tilde{r}_A}\,.
\eea
Because of the spherical symmetry of the Schwarzschild spacetime, we can always adjust the coordinates so the geodesic lies entirely on the $\tilde{\theta}_A= \pi/2$ plane (see Fig.~\ref{fig:Zones}). The symmetries (including the metric being temporally static) also allows for the identification of two conserved quantities, the specific (per unit rest mass) energy $\mathcal{E}$ and specific angular momentum $\mathcal{L}$, defined by 
\bea
\mathcal{E}= f(\tilde{r}_A)\frac{d\tilde{t}_A}{d\tau}\,, \quad \mathcal{L} = \tilde{r}^2_A \frac{d\tilde{\phi}_A}{d\tau}\,, 
\eea
as well as an impact-parameter-like length scale $\beta \equiv \mathcal{L}/\mathcal{E}$, allowing us to manipulate Eq.~\eqref{eq:SchMetric} into 
\begin{align}\label{eq:SchGeoEq}
\left( \frac{d\tilde{u}_A}{d\tilde{\phi}_A}\right)^2 =& \frac{1}{
\beta^2} - \left(1-2M_A\tilde{u}_A\right)\left(\frac{1}{\mathcal{L}^2}+\tilde{u}^2_A\right)
\notag \\
\equiv& 2M_A(\tilde{u}_A-u_1)(\tilde{u}_A-u_2)(\tilde{u}_A-u_3)
\,,
\end{align}
where in the second line we factorize the cubic polynomial in terms of the inverse radius $\tilde{u}_A \equiv 1/\tilde{r}_A$. Solution to Eq.~\eqref{eq:SchGeoEq} is 
\begin{align} \label{eq:SchGeoSol}
\tilde{u}_A = u_1 + \left(u_2-u_1\right)\text{sn}^2\left( \frac{1}{2} \tilde{\phi}_A\sqrt{2M_A (u_3-u_1)}+ \phi_0\,\bigg| \frac{u_2-u_1}{u_3-u_1}\right)\,,
\end{align}
where $\text{sn}$ is the Jacobi elliptic sine function. The parameter $\phi_0$ helps specify the initial angular position, against the $\tilde{\phi}_A=0$ axis, which for our purpose is most conveniently set according to the direction of the BH's linear motion. The explicit expressions for the roots are 
\begin{align} \label{eq:ExplicitU}
u_1 =&\frac{1}{6 M_A}\left( 1-\frac{(1+i\sqrt{3})\beta^2\xi}{2\zeta}-\frac{(1-i\sqrt{3})\zeta}{2\beta^2 \mathcal{L}^2}\right)\,,\notag \\
u_2 =&\frac{1}{6 M_A}\left( 1-\frac{(1-i\sqrt{3})\beta^2\xi}{2\zeta}-\frac{(1+i\sqrt{3})\zeta}{2\beta^2 \mathcal{L}^2}\right)\,,\notag \\
u_3 =& \frac{1}{6 M_A}\left( 1+\frac{\beta^2\xi}{\zeta}+\frac{\zeta}{\beta^2 \mathcal{L}^2}\right) \,,
\end{align}
where 
\begin{align}
\zeta \equiv& \Bigg\{ \beta^6 \mathcal{L}^4\left(\mathcal{L}^2+36 M_A^2\right)-54 \beta^4 \mathcal{L}^6 M_A^2+6 \sqrt{3} \sigma^{1/2}\Bigg\}^{1/3}\,, \notag \\
\sigma =& \beta^8 \mathcal{L}^6 M_A^2 \bigg(\beta^4 \left(\mathcal{L}^2+4 M_A^2\right)^2
-\beta^2 \mathcal{L}^4\left(\mathcal{L}^2+36 M_A^2\right)+27 \mathcal{L}^6 M_A^2\bigg)\,,\notag \\
\xi \equiv& \mathcal{L}^2-12M_A^2 \,. 
\end{align}
Since the three roots must add up to $1/(2M_A)$, and $u_3 \in \mathbb{R}$, the roots $u_1$ and $u_2$ are either (i) complex conjugations of each other, or (ii) both real (no need to be equal in this case). Since $\xi \in \mathbb{R}$ and $\zeta \in \mathbb{C}$, we have that 
\bea
\bar{u}_2 =\frac{1}{6 M_A}\left( 1-\frac{(1+i\sqrt{3})\beta^2\xi}{2\bar{\zeta}}-\frac{(1-i\sqrt{3})\bar{\zeta}}{2\beta^2 \mathcal{L}^2}\right)\,. 
\eea
Comparing with $u_1$ in Eq.~\eqref{eq:ExplicitU}, we see that case (i) generically happens when $\sigma > 0$ (so $\zeta$ could be real), while (ii) happens when $\sigma < 0$, or in other words when the impact parameter is sufficiently large
\bea
\beta > \frac{\mathcal{L}\sqrt{\mathcal{L} \left(\mathcal{L}^3+36 \mathcal{L}M_A^2 - \xi^{3/2}\right)}}{\sqrt{2}(\mathcal{L}^2 + 4M_A^2)}\,, 
\eea
or in terms of $\mathcal{E}$, that
\bea
\mathcal{L} > \mathcal{L}^{\rm min} \equiv \frac{M_A\sqrt{27\mathcal{E}^4-36\mathcal{E}^2+8+\mathcal{E}(9\mathcal{E}^2-8)^{3/2}}}{\sqrt{2(\mathcal{E}^2-1)}}\,. 
\eea
The hyperbolic trajectories in Newtonian gravity generalizes (some may wind around the BH before escaping, see Fig.~\ref{fig:GeoEndsPics}) to geodesics in category (ii), and it is these geodesics that are relevant to our slingshots. For them, the value $1/u_2$ gives the distance of closest approach (the roots are places where $d\tilde{u}_A/d\tilde{\phi}_A=0$, thus correspond to extrema or inflection points in $\tilde{r}_A$). 
For completeness, we note that those other trajectories with smaller $\mathcal{L}$ mostly end up falling into the BH. A very convenient but rough way to see this is by considering the radial motion only, which satisfies 
\begin{align} \label{eq:GRPot}
\left(\frac{d\tilde{r}_A}{d\tau}\right)^2 &= \mathcal{E}^2 - V(\tilde{r}_A)\,, \notag \\
V(\tilde{r}_A) &\equiv f(\tilde{r}_A)+\frac{\mathcal{L}^2}{\tilde{r}^2_A}-\frac{2M_A \mathcal{L}^2}{\tilde{r}^3_A}\,,
\end{align}
and resembles a particle of unit mass and energy $\mathcal{E}^2/2$ moving in a one dimensional potential well $V(\tilde{r}_A)/2$ that is pictorially depicted in Fig.~\ref{fig:Potential}. It is clear then that for a given $\mathcal{E}$, having too small an $\mathcal{L}$ would allow the particle to slip into the BH through the ``central valley'' beneath the $\mathcal{E}^2$ surface. 

\begin{figure}
  \centering
\begin{overpic}[width=0.49\columnwidth]  {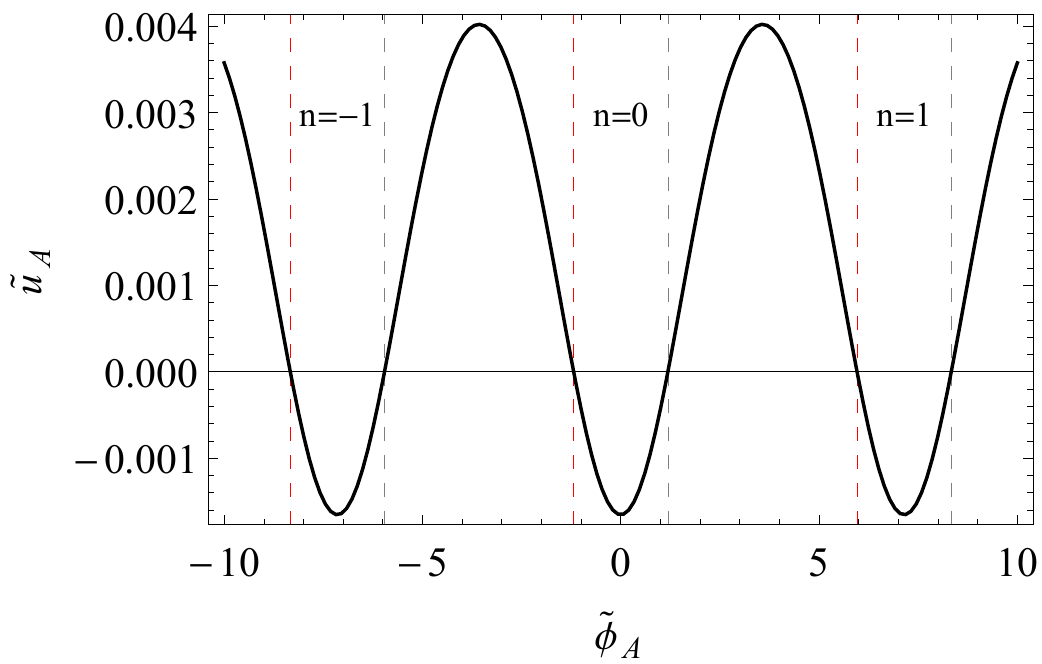}
\put(5,0){(a)}
\end{overpic}\vspace{3mm}
\begin{overpic}[width=0.49\columnwidth]  {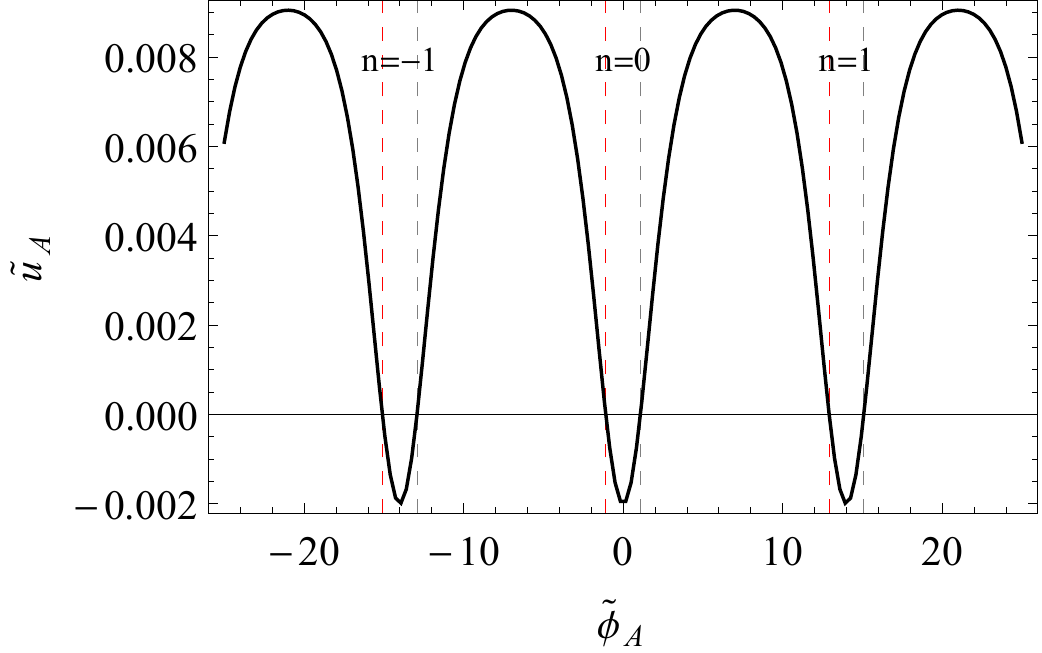}
\put(5,0){(b)}
\end{overpic}
  \caption{The inverse-radius $\tilde{u}_A$ as a function of $\tilde{\phi}_A$, corresponding to the two cases of Fig.~\ref{fig:GeoEndsPics}. The red/black dashed vertical lines are $\tilde{\phi}^+_A/\tilde{\phi}^-_A$ with the corresponding $n$ numbers written next to them. Only the $\tilde{u}_A>0$ parts correspond to physical geodesics, and the different segments separated by $\tilde{u}_A<0$ ditches are equivalent geodesics shifted in $\tilde{\phi}_A$, a freedom that is already accounted for by $\phi_0$, so we only need to consider one segment. (a): For a more traditional ``hyperbolic'' orbit with $\mathcal{L}$ much larger than $\mathcal{L}^{\rm min}$. (b): For a geodesic with $\mathcal{L}$ only slightly larger than $\mathcal{L}^{\rm min}$, which winds around the BH for a full cycle before escaping. }
	\label{fig:GeoEnds}
\end{figure}

Having now the correct geodesics, Eq.~\eqref{eq:SchGeoSol} then allows to compute $d\tilde{u}_A/d\tilde{\phi}_A$ on such trajectories, and so we can obtain the nonvanishing components of the proper velocity via 
\bea 
\frac{d\tilde{\phi}_A}{d\tau} = \frac{\mathcal{L}}{\tilde{r}^2_A}\,, \quad 
\frac{d\tilde{r}_A}{d\tau} = -\tilde{r}^2_A\frac{d\tilde{u}_A}{d\tilde{\phi}_A}\frac{d\tilde{\phi}_A}{d\tau} = -\mathcal{L}\frac{d\tilde{u}_A}{d\tilde{\phi}_A}\,, 
\eea
and then the comoving coordinate velocity is given by 
\begin{align} \label{eq:ComovingVelo}
\frac{d\tilde{\phi}_A}{d\tilde{t}_A} &= \frac{d\tilde{\phi}_A}{d\tau}\frac{d\tau}{d\tilde{t}_A} = \beta\frac{f(\tilde{r}_A)}{\tilde{r}^2_A}\,, \quad \frac{d\tilde{\theta}_A}{d\tilde{t}_A} =0\,, \notag \\ 
\frac{d\tilde{r}_A}{d\tilde{t}_A} &= \frac{d\tilde{r}_A}{d\tau}\frac{d\tau}{d\tilde{t}_A} = -\beta \,f(\tilde{r}_A) \frac{d\tilde{u}_A}{d\tilde{\phi}_A}\,.
\end{align}
Because $\tilde{u}_A$, thus $\tilde{r}_A$ and $d\tilde{u}_A/d\tilde{\phi}_A$ are all functions of $\tilde{\phi}_A$, these velocity components are most conveniently expressed in this variable (no need to integrate $d\tilde{\phi}_A/d\tilde{t}_A$ and re-express everything as functions of $\tilde{t}_A$), with the initial $\tilde{\phi}^{\iota}_A$ and ending $\tilde{\phi}^{\epsilon}_A$ values given by the solutions of $\tilde{u}_A=0$ ($\tilde{r}_A \rightarrow \infty$) according to Eq.~\eqref{eq:SchGeoSol}.  
Explicitly, these asymptotic $\tilde{\phi}_A$ values are 
\begin{align} \label{eq:Asymptotes}
\tilde{\phi}^{\pm}_A =& -\frac{\sqrt{2}}{\sqrt{M_A (u_3-u_1)}} \Bigg[\phi_0\pm \text{sn}^{-1}\left(\frac{\sqrt{u_1}}{\sqrt{u_1-u_2}}\Bigg|\frac{u_2-u_1}{u_3-u_1}\right)
+ 2 n \mathcal{K}\left(\frac{u_2-u_1}{u_3-u_1}\right)\Bigg]\,, \quad n\in \mathbb{Z}\,,
\end{align}
where $\mathcal{K}$ is the complete elliptic integral of the first kind. We plot the function $\tilde{u}_A(\tilde{\phi}_A)$ in Fig.~\ref{fig:GeoEnds}, which makes it clear that we can pick 
\bea \label{eq:Asymptotes2}
\tilde{\phi}^{\iota}_A = \tilde{\phi}_A^-(n=0)\,,\quad \tilde{\phi}^{\epsilon}_A = \tilde{\phi}_A^+(n=1)\,,
\eea
or vice versa (in the comoving frame the geodesics are time reversal symmetric). Note further that there are special values for the free parameter $\phi_0$ as well, since when 
\bea
\phi_0 = \phi_0^{\rm sym+} \equiv - \mathcal{K}\left(\frac{u_2-u_1}{u_3-u_1}\right)\,,
\eea
we have $\tilde{\phi}_A^{\iota} = -\tilde{\phi}_A^{\epsilon}$, or in other words the symmetric case where the BH velocity sits midway between the incoming and outgoing directions. This symmetry is of course also true if ${\bf v}_A$ points in the opposite direction, corresponding to 
\bea
\phi_0 = \phi_0^{\rm sym-} \equiv \phi_0^{\rm sym+} + \frac{\pi \sqrt{M_A (u_3-u_1)}}{\sqrt{2}}\,.
\eea 
 
\subsubsection{Gravity assist} \label{sec:Assist}
In order to compute the speed gain picked up by particles following the trajectories described in the last section, we need to change into a ``resting'' reference frame where the BH is moving. Instead of going into the binary frame (at the centre of Fig.~\ref{fig:Zones}) though, it is less messy and thus more illuminating to utilize an intermediate frame (whose associated quantities are denoted with an overhead hat) where the BH is at the origin, but moving with a linear velocity matching its instantaneous velocity ${\bf v}_A$ in the binary frame. 
We also align the axes of the intermediate frame with those of the comoving frame, so the geodesic is on the $\hat{\theta}=\pi/2$ plane. In other words, the intermediate frame is translated and rotated against the binary frame, but is instantaneously stationary against it (over much longer time scales than that of a slingshot, the intermediate frame rotates against the binary frame, please refer to discussions at the end of Sec.~\ref{sec:SingleSling}), thus would record the same particle speeds. 
Essentially, we are borrowing the technique commonly employed for constructing numerical initial data for BBH simulations (see e.g., \cite{2009CQGra..26k4002L}), namely approximating the metric near the BHs (where slingshots take place) by linearly boosted single hole solutions\footnote{Which are legitimate asymptotically manifestly Minkowski exact solutions to general relativity that contain moving BHs, constructed by applying a global Lorentz coordinate transformation to the stationary BH solutions, even though the spacetime is not Minkowski everywhere.}. Such strategies have been shown to produce sensible results, in that the simulation quickly settles into a quasi-circular binary motion with minimal changes to BH parameters (such as spin) or binary parameters (such as separation), which is only possible is the initial data is already a fairly good approximate.

It is then easy to apply the Lorentz velocity transformation formula
\bea \label{eq:BoostVelo}
\hat{\bf v}=\frac{1}{1+\tilde{\bf v}\cdot {\bf v}_A} \left(
\frac{\tilde{\bf v}}{\gamma}+{\bf v}_A+\frac{\gamma}{1+\gamma}\left(\tilde{\bf v}\cdot {\bf v}_A\right){\bf v}_A
\right)\,,
\eea
where $-{\bf v}_A$ is the velocity of the intermediate frame as measured in the comoving frame, and $\gamma = 1/\sqrt{1-|{\bf v}_A|^2}$ is the associated Lorentz factor.
One cautionary note though is that we are not dealing with a Minkowski metric when close to the BH, so while it is obviously easier to apply Eq.~\eqref{eq:BoostVelo} when both the comoving and the intermediate frame quantities are expressed under Cartesian coordinates, one needs to be careful when constructing these coordinates. The standard rules of 
\begin{align} \label{eq:CoordSpToCart}
\tilde{x}_A &= \tilde{r}_A\sin\tilde{\theta}_A\cos\tilde{\phi}_A\,, 
\quad 
\tilde{y}_A = \tilde{r}_A\sin\tilde{\theta}_A\sin\tilde{\phi}_A\,, 
\quad 
\tilde{z}_A = \tilde{r}_A\cos\tilde{\theta}_A\,, 
\end{align}
do not lead to orthogonal Cartesian basis vectors under metric \eqref{eq:SchMetric} if $f(\tilde{r}_A)\neq 1$, and so the process of taking projections becomes more complicated. Fortunately though, we are only interested in the asymptotic speeds of the particle when $\tilde{r}_A \rightarrow \infty$, so we only need to deal with the flat regions and Eq.~\eqref{eq:CoordSpToCart} can be applied without complications, together with a number of additional simplifications (we will use the subscript $\cdot_{\infty}$ to flag expressions only valid in this context).  
To begin with then, recall that we can chosen the $\tilde{\phi}_A=0$ or $\tilde{x}_A$ direction by utilizing the freedom endowed by $\phi_0$, and we find it convenient to fix it to be the projection $\mathbb{P}({\bf v}_A)$ of ${\bf v}_A$ onto the plane of the geodesic (see Fig.~\ref{fig:Zones}). Let $\tilde{\Theta}_A$ be the angle between ${\bf v}_A$ and the $\tilde{z}_A$ axis in the comoving frame, then from Eq.~\eqref{eq:ComovingVelo}, we have 
\bea
\mathcal{N}_{\infty}\equiv \tilde{\bf v}\cdot {\bf v}_A\bigg|_{\tilde{r}_A \rightarrow \infty} =  -\beta |{\bf v}_A| \sin\tilde{\Theta}_A \cos \tilde{\phi}_A \frac{d\tilde{u}_A}{d\tilde{\phi}_A}\,.
\eea
Note that although we are in the intermediate frame, we can still use $\tilde{\phi}_A$ as an abstract parameter along the geodesic. We have then that 
\begin{align} \label{eq:InterFrame}
\hat{v}_{\infty}^{\hat{x}_A} =& \frac{1}{\mathcal{N}_{\infty}+1}
\Bigg[-\frac{\beta }{\gamma }\frac{d\tilde{u}_A}{d\tilde{\phi}_A}  \cos \tilde{\phi}_A
+\left( \frac{\gamma + \gamma  \mathcal{N}_{\infty} + 1}{\gamma +1}\right)|{\bf v}_A| \sin \tilde{\Theta}_A\Bigg]\,, \notag \\
\hat{v}_{\infty}^{\hat{y}_A} = & -\frac{\beta }{\gamma   (\mathcal{N}_{\infty}+1)} \frac{d\tilde{u}_A}{d\tilde{\phi}_A} \sin \tilde{\phi}_A\,, \notag \\
\hat{v}_{\infty}^{\hat{z}_A} = & \left(\frac{\gamma +\gamma  \mathcal{N}_{\infty}+1}{\gamma +1}\right)\frac{|{\bf v}_A| \cos \tilde{\Theta}_A }{\mathcal{N}_{\infty}+1}\,,  
\end{align}
where explicitly 
\begin{align}
\frac{d\tilde{u}_A}{d\tilde{\phi}_A} =& \frac{i \Sigma  \left(\beta^4 \mathcal{L}^2 \xi -\zeta ^2\right)}{6 \sqrt{2} \beta^2 \zeta  \mathcal{L}^2 M_A} \text{cn}\left(\left.\frac{\tilde{\phi}_A \Sigma }{2 \sqrt{6}}+\phi_0\right|\Pi \right)
\text{dn}\left(\left.\frac{\tilde{\phi}_A \Sigma }{2 \sqrt{6}}+\phi_0\right|\Pi \right) 
 \text{sn}\left(\left.\frac{\tilde{\phi}_A \Sigma }{2 \sqrt{6}}+\phi_0\right|\Pi \right)\,,
\end{align}
with cn and dn being the Jacobi elliptic cosine and delta functions, and 
\begin{align}
\Sigma =& \sqrt{\frac{\left(3+i \sqrt{3}\right) \beta^2 \xi }{\zeta }+\frac{\left(3-i \sqrt{3}\right) \zeta }{\beta^2 \mathcal{L}^2}}\,, \notag \\
\Pi =& \left(\frac{\sqrt{3} i \beta^4 \mathcal{L}^2 \xi }{\zeta ^2-\beta^4 \mathcal{L}^2 \xi }+\frac{\sqrt{3} i}{2}+\frac{1}{2}\right)^{-1}\,.
\end{align}
We have now all the ingredients for computing the initial and ending speeds $|\hat{\bf v}_{\infty}^{\iota/\epsilon}|\equiv|\hat{\bf v}(\tilde{\phi}^{\iota/\epsilon})|$ and Lorentz factors
\bea
\gamma^{\iota/\epsilon} \equiv \frac{1}{\sqrt{1-\big|\hat{\bf v}^{\iota/\epsilon}_{\infty}\big|^2}}\,,
\eea
allowing us to evaluate the effectiveness of slingshots in the relativistic context. Strictly speaking, $\tilde{\phi}^{\iota/\epsilon}$ are asymptotic values, and the initial and ending speeds need to be obtained via a limit-taking process. However, the velocity of the particle will not change much once it is far away from the BH, so the limit can be obtained simply by plugging the $\tilde{\phi}^{\iota/\epsilon}$ values given by Eqs.~\eqref{eq:Asymptotes} and \eqref{eq:Asymptotes2} into the $\hat{\bf v}_{\infty}$ expression (unlike $\tilde{r}_A$, there are no divergence issues with $\hat{\bf v}$ there). 

\subsubsection{Parameter dependence}\label{sec:Params}
\begin{figure}
  \centering
\begin{overpic}[width=0.49\columnwidth]  {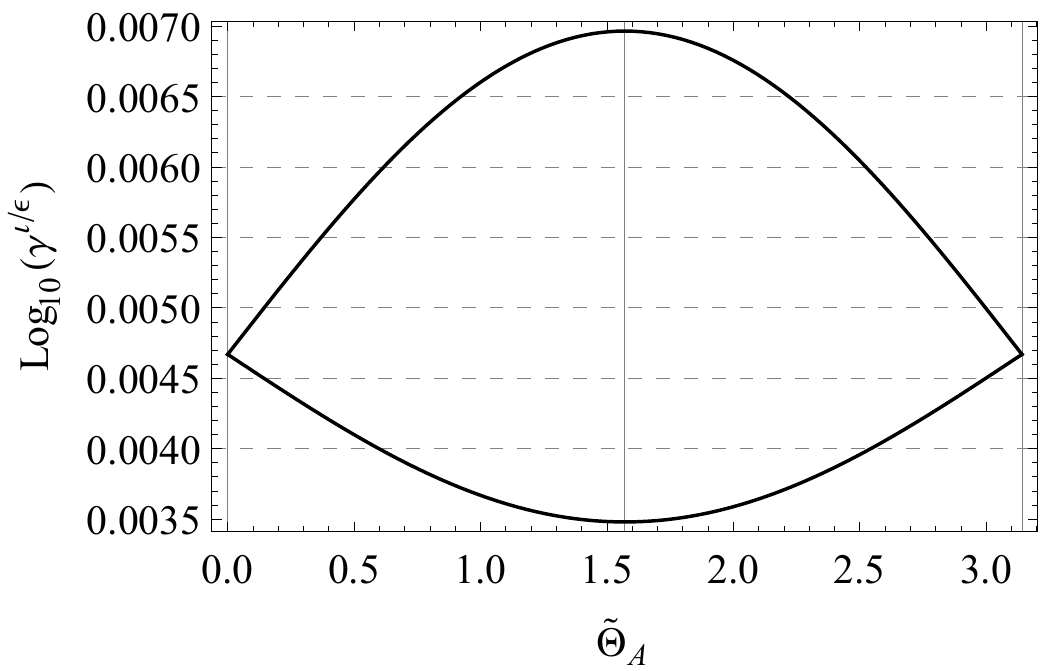}
\put(5,5){(a)}
\end{overpic}\vspace{3mm}
\begin{overpic}[width=0.49\columnwidth]  {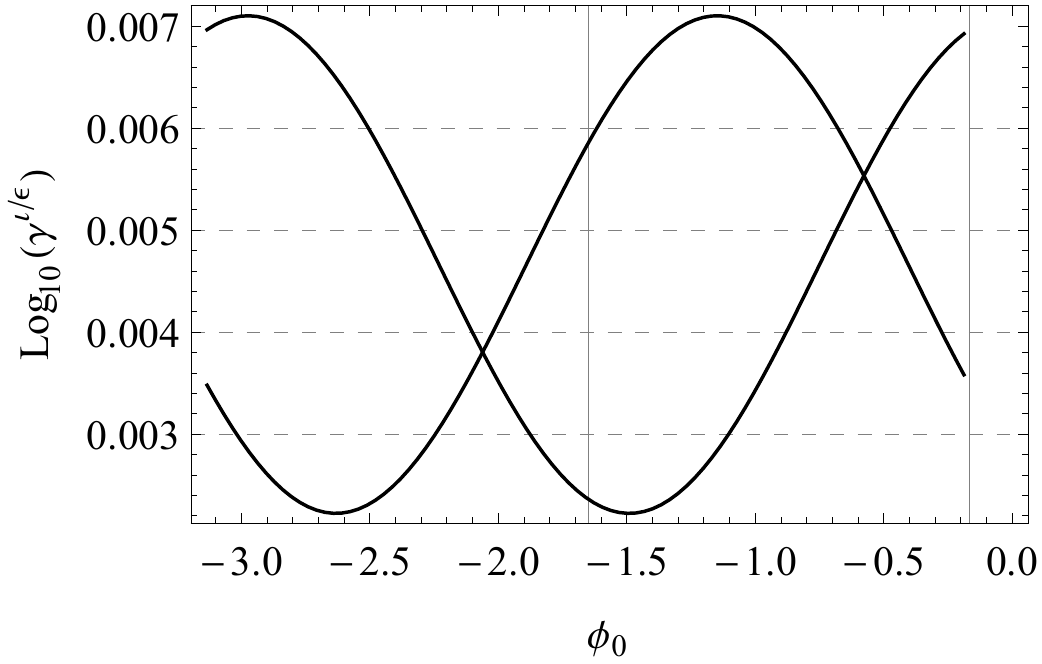}
\put(5,5){(b)}
\end{overpic}
  \caption{
   (a): Plot of $\gamma^{\iota/\epsilon}$ as functions of $\tilde{\Theta}_A$, with $\phi_0=\phi_0^{\rm sym-}$. The solid vertical lines mark out $\tilde{\Theta}_A=0$, $\pi/2$ and $\pi$. 
(b): Plot of $\gamma^{\iota/\epsilon}$ as functions of $\phi_0$, with $\tilde{\Theta}_A=\pi/2$. The two solid vertical lines correspond to $\phi_0^{\rm sym+}$ and $\phi_0^{\rm sym-}$. 
The other parameters are fixed at $\mathcal{E}=1.01$, $\mathcal{L}=1.5\mathcal{L}^{\rm min}$ and $|{\bf v}_A|=0.04$c for both panels. }
	\label{fig:ThetaDep}
\end{figure}

To gauge the efficiency of the relativistic slingshots and their parameter dependence, we compute some concrete example speed/Lorentz factor gains with representative configurations. 
Overall, we have two adjustable BH parameters $M_A$ and $|{\bf v}_A|$, and four geodesic parameters $\phi_0$, $\tilde{\Theta}_A$, $\mathcal{E}$ and $\mathcal{L}$ (noticeably but not surprisingly absent is the mass of the particle, so all nuclei will be accelerated to the same Lorentz factor). The parameter space is thus too high in dimensions for us to explore exhaustively. Instead, we restrict the more constrained BBH parameters to a limited set of fiducial values. Specifically, following earlier discussion below Eq.~\eqref{eq:Kepler} (see also discussion later in Sec.~\ref{sec:EvRate}), we continue to use the example value $|{\bf v}_A| \sim 0.01$c (corresponding to $b\sim 2000M_{\rm tot}$ by Eq.~\eqref{eq:Kepler}), but also $0.04$c ($b\sim 200M_{\rm tot}$), $0.1c$ ($b\sim 20M_{\rm tot}$) and $0.2c$ ($b\sim 6M_{\rm tot}$), to show scaling against $|{\bf v}_A|$. We also set $M_A = 30 \text{M}_{\odot}$, roughly matching the first binary seen in GWs; but note that changing this value won't alter any results in this section, since it is the only length scale in the problem, and speeds are dimensionless in geometrized units (any two quantities whose ratio is a speed must have the same dimensions, and thus scale with $M_A$ in the same way  -- since it is the only length scale, leading to the scaling factors cancelling in the ratio). 
The parameter $\tilde{\Theta}_A$ is also not particularly interesting. Having ${\bf v}_A$ outside of the plane of the geodesic simply reduces the effectiveness of the gravity assist, as Fig.~\ref{fig:ThetaDep}(a) demonstrates. Since we are interested in the maximum boost in Lorentz factors achievable, we set $\tilde{\Theta}_A = \pi/2$ throughout the remainder of this section. 

\begin{figure*}
  \centering
\begin{overpic}[width=0.45\textwidth]  {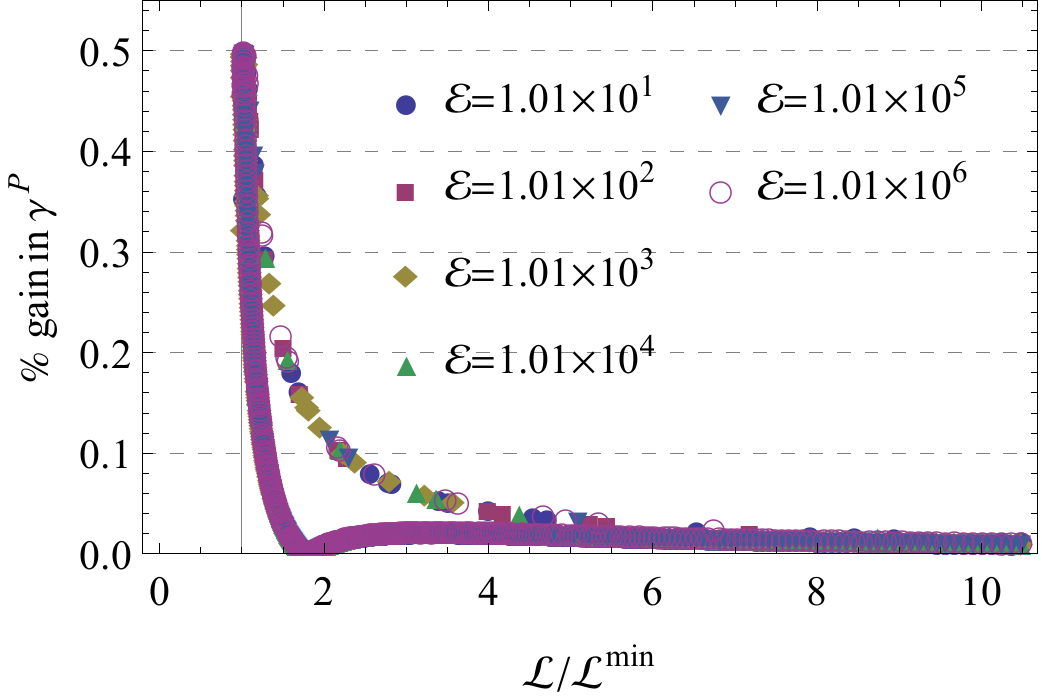}
\put(5,0){(a)}
\end{overpic}
\begin{overpic}[width=0.45\textwidth]  {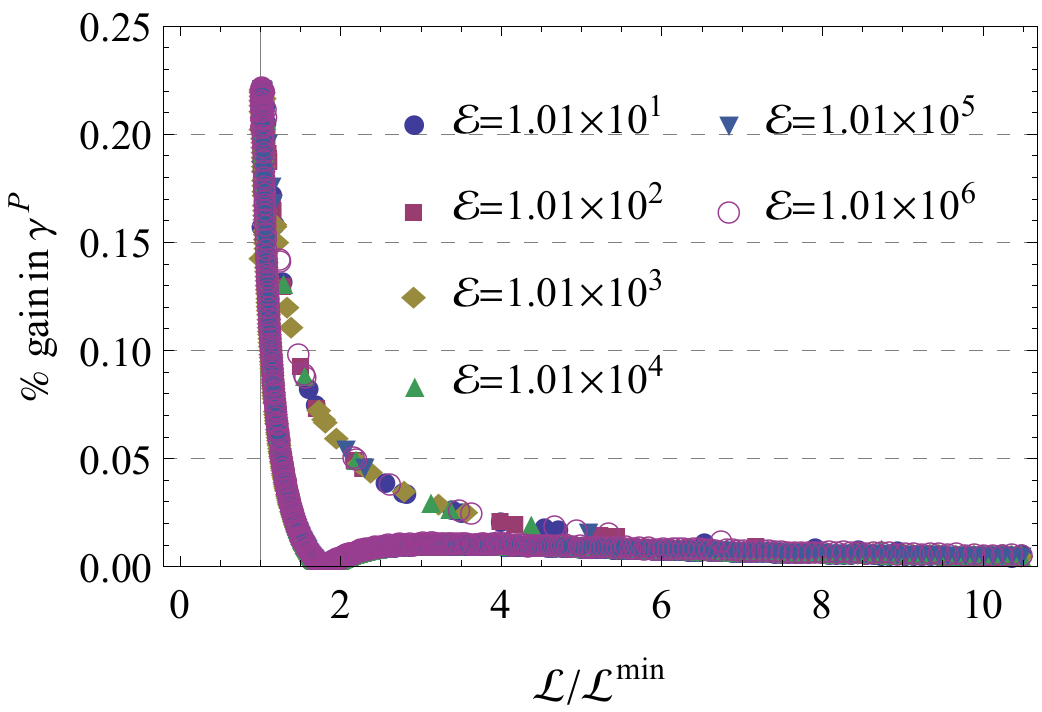}
\put(5,0){(b)}
\end{overpic}\vspace{3mm}
\begin{overpic}[width=0.45\textwidth]  {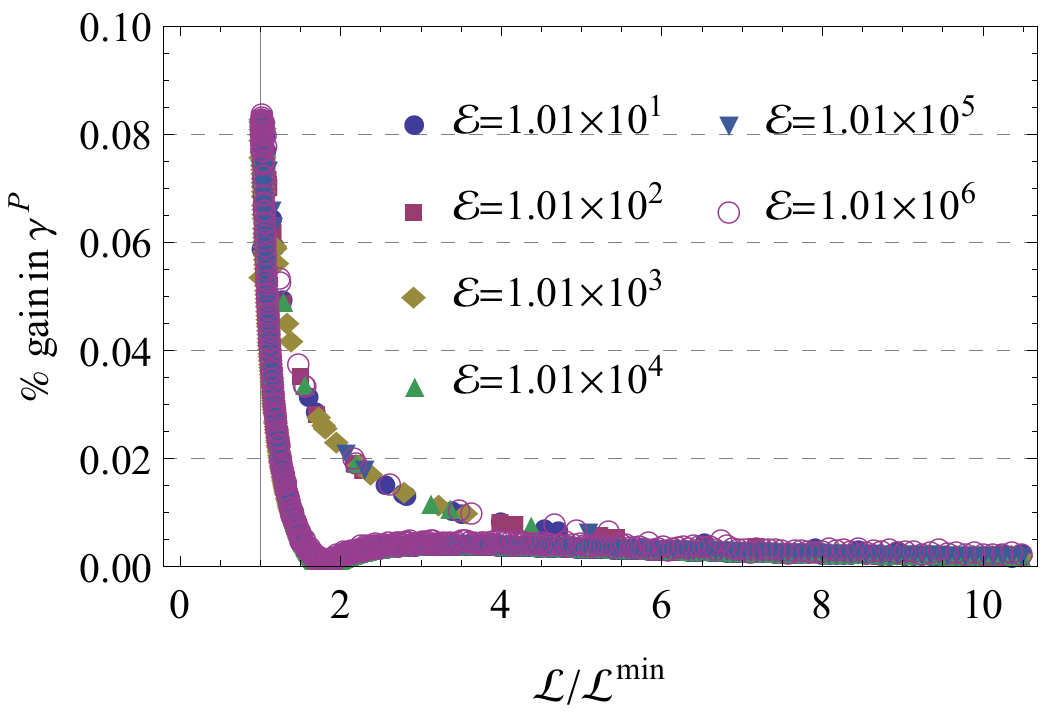}
\put(5,0){(c)}
\end{overpic}\vspace{3mm}
\begin{overpic}[width=0.45\textwidth]  {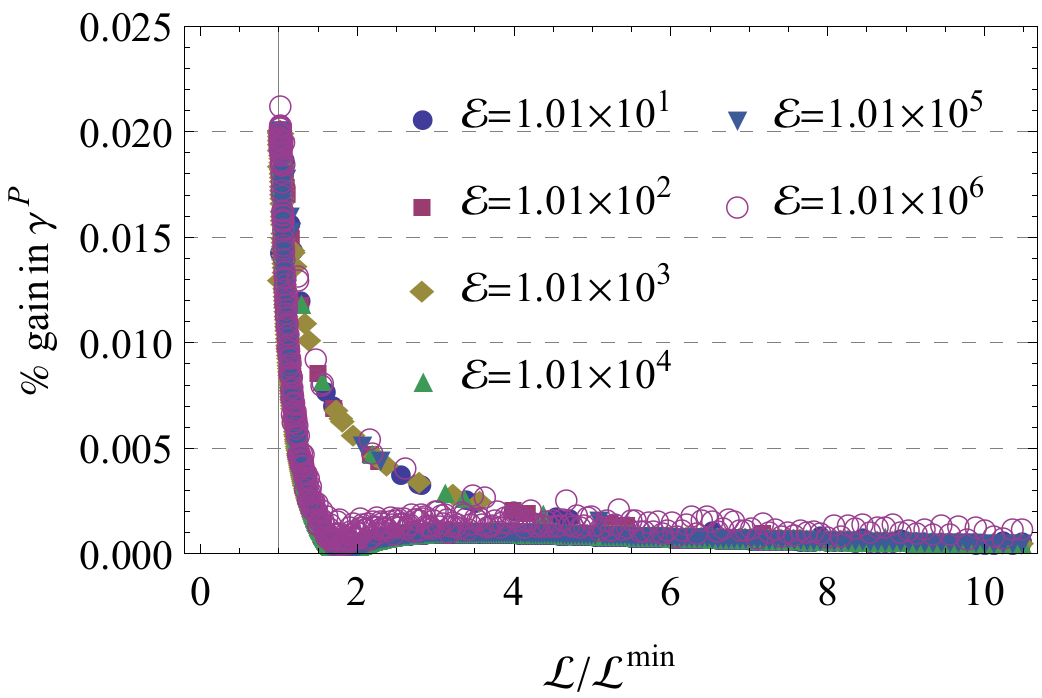}
\put(5,0){(d)}
\end{overpic}
  \caption{The percentage (e.g., $0.2$ on the vertical axis means $20\%$) gain in the Lorentz factor $\gamma^P$ of a particle through an episode of gravitational slingshot, plotted as a function of normalized specific angular momentum. Each data point is optimized over $\phi_0$. Panel (a) is for $|{\bf v}_A|=0.2$c, (b) for $|{\bf v}_A|=0.1$c, (c) for $|{\bf v}_A|=0.04$c and (d) for $|{\bf v}_A|=0.01$c, note the accuracy of the numerics degrades when both very large ($\mathcal{E}$) and very small ($|{\bf v}_A|$) numbers appear simultaneously. Within each panel, there are two branches, with the lower one densely filled (all $\mathcal{E}$ values are present, but their points overlap so only the marker of one of them is obvious), and the upper one sparsely populated. Geodesics landing on the upper branch are thus comparatively rare and less important. 
}
	\label{fig:Speedup}
\end{figure*}

The most interesting and variable parameter is the specific energy $\mathcal{E}$, which asymptotically ($f(\tilde{r}_A)\approx 1$) is the particle's Lorentz factor in the BH's comoving frame -- not the binary or intermediate frame (i.e., not the specific energy we measure for the UHECRs on Earth, otherwise there won't be a speed-up due to slingshots, since $\mathcal{E}$ is conserved along geodesics). It is however not too far removed from the binary frame value (because the BHs' speeds are not quite $\mathcal{O}(1)c$, so the differences between the two frames are not exceedingly high), and can be used as a rough indicator of the latter. For application to UHECR, it is important to examine how the speed gain per slingshot evolves when the particle has already sped up during previous episodes, i.e., how it evolves against increases in $\mathcal{E}$ (note that $\mathcal{E}$ is only conserved during each individual slingshot, and not when the geodesic is threading through the binary, whose metric is not stationary; so subsequent slingshots will carry increased $\mathcal{E}$, reflecting the gains from previous slingshots). With each $\mathcal{E}$ value, it is also interesting to know the speed-up's dependence on how close the geodesics get to the BHs (i.e., on the $\mathcal{L}$ values). 
We therefore numerically compute, for each $\mathcal{E}$ choice, the maximum percentage Lorentz factor gain as optimized over $\phi_0$ (see Fig.~\ref{fig:ThetaDep}(b) for dependence of $\gamma^{\iota/\epsilon}$ on $\phi_0$), as a function of the ratio $\mathcal{L}/\mathcal{L}^{\rm min}$. The results of this exercise is presented in Fig.~\ref{fig:Speedup}.

Naively, one might worry that since slingshots cannot possibly speed up a particle already moving at the speed of light, their efficiency would diminish as a particle speeds up towards c. While this is certainly true in terms of speed gains, Fig.~\ref{fig:Speedup} shows that the situation is different with the Lorentz factor or specific energy (which increases more rapidly against speed as the latter becomes larger) that we care about. From the figure, we immediately notice a universality in terms of $\mathcal{E}$, indicating that as the particle energizes up through successive slingshot instances, the efficiency of each subsequent slingshot does not decline or increase. 
Furthermore, the relative gain scales roughly (slightly more aggressively than) linearly with $|{\bf v}_A|$. 

\subsection{Repeated slingshots}
Fig.~\ref{fig:Speedup} demonstrates that within the allowed parameter range, it is impossible to catapult particles to the UHECR energies with a single slingshot, so multiple slingshots will be necessary. It is obviously too restrictive to fine-tune the initial conditions of the geodesics so that all the slingshots accelerate particles with nearly maximal gain. Barring the existence of attractor resonant orbits, such configurations would be unrealistic to achieve in sufficient prevalence to reproduce the UHECR luminosity within real astrophysical settings. Instead, we envisage that the relevant geodesics to be quite arbitrary and chaotic (see Fig.~\ref{fig:Trajs}), containing not only sub-optimal accelerating slingshots but also decelerating ones, making the RS acceleration mechanism essentially a statistical one, as opposed to a direct one. Just like the Fermi process then, the accelerating slingshots must somehow win out over the decelerating ones. 

This is indeed the case. Because of aberration, a cloud of particles coming towards a BH in the binary frame from all angles in a uniform distribution would crowd onto the head-on direction in the BH's comoving frame (where the spherical symmetry allows these geodesics to be treated on an equal footing; so this frame is appropriate for statistics), meaning that with particles moving about rather randomly in the binary frame, it is more likely to have them scatter against the BHs in a roughly head-on rather than head-tail configuration, implying that acceleration rather than deceleration would win out statistically. Alternatively, one may simply borrow Fermi's argument that head-ons are more likely due to greater relative velocities \cite{1949PhRv...75.1169F}, which allows for an estimate of the relative probabilities of accelerating versus decelerating events occurring (proportional to the relative velocities, see discussion below his Eq.~13), which then raises the average (weighted according to said probabilities) percentage energy gain's dependence on $v_{\rm BH}$ from linear (when only accelerating slingshots occurs along the UHECR geodesics) to quadratic. In other words, the RS mechanism is similar to the original Fermi acceleration, i.e., being of second order. Fortunately though, the scattering centres in RS are BHs with relativistic speeds (while those in the Fermi acceleration are slow moving magnetic perturbations), so the loss of efficiency is nowhere near as severe (i.e., unlike the Fermi process, there is not as much incentive to elevate the RS process to first order).

For our study, it is also useful to consider hypothetical optimal orbits that can achieve UHECR energies with an absolute minimal number of slingshots. This is because, first of all, they provide an indication as to by how much estimates based on Newtonian elastic scattering formulae (see Sec.~\ref{sec:Newtonian}), ubiquitously adopted in less extreme slingshot contexts, would have been off, thereby justifying our extra effort in deriving the relativistic expressions. They can also serve as ``standard reference cases'' that can be used to compare the RS properties across different binary configurations. Specifically, the simple near-linear dependence on BH speed $|{\bf v}_A|$ allows for a very simple estimate on the minimal number $N_{\rm min}$ of slingshots required to reach UHECR energies. Take the frequently evoked typical UHECR energy of $10^{19}$eV, translating into a Lorentz factor of around $10^{9}$ for nitrogen nuclei ($14$ nucleon with rest mass $\approx 1$GeV each), then in the upper limit case of a $50\%$ gain when $|{\bf v}_A|=0.2$c, we have that 
\bea \label{eq:NoShots}
N^{0.2k c}_{\rm min}=\log_{1+0.5k}\left(\frac{10^9}{\gamma^P_0}\right)\,.
\eea 
Note that when arriving at Eq.~\eqref{eq:NoShots}, not only have we used the maximum gain across all possible $\mathcal{L}/\mathcal{L}^{\rm min}$ (after already having optimized over $\phi_0$ and $\tilde{\Theta}_A$). To evaluate these $N^{0.2k c}_{\rm min}$ into concrete numbers, we need to estimate the initial Lorentz factor $\gamma^P_0$. Note that when the particles first become captured by the binary, they would have lower but still relativistic speeds, since they would have to fall into the NZ of the binary before the slingshot actions can ensure. Assuming an initially stationary particle being attracted gravitationally towards the binary, falling from infinity into the NZ with boundary at $R_{\rm NZ}\sim 20000M_{\rm tot}$, then equating the specific gravitational potential $M_{\rm tot}/R_{\rm NZ}$ to the specific kinetic energy gain of $\gamma^P_0-1$\footnote{There will be the same amount of energy lost when the particles eventually escape to Earth, but as a percentage of the ultra-high energy carried by the particles then, this effect is negligible.}, we obtain $\gamma^P_0 \sim 1.00005$ (or an initial speed of $0.01$c),
which gives 
\begin{align}\label{eq:NoSlings}
N^{0.2c}_{\rm min} \approx 50\,, \quad N^{0.1c}_{\rm min}\approx 100\,, \quad N^{0.04c}_{\rm min} \approx 250\,, \quad N^{0.01c}_{\rm min} \approx 1000\,. 
\end{align} 
This estimate is of course subject to uncertainties. For example, due to technical difficulties, we have not considered rotating BHs, but a fast-spinning BH can plausibly further fling the particles as they make close passes near its event horizon (i.e., increase the gain per slingshot). Furthermore, as already alluded to in Sec.~\ref{sec:SlingsIntro}, a rapidly spinning BH will help prevent particles from falling into it. Because the GW data \cite{2016PhRvL.116x1103A,2017PhRvL.118v1101A,2017ApJ...851L..35A,2017PhRvL.119n1101A} thus far have appeared consistent with moderate individual BH spins however, we will not pursue this matter further.

The differences in $N_{\rm min}$ as given by Eq.~\eqref{eq:NoSlings} for the different stages of the binary inspiral has important implications for deciding which stage of the BBH coalescence is the most relevant in terms of UHECR production. These minimal values are already larger than the number of deflection events seen for the arbitrary (generic) trajectories displayed in Fig.~\ref{fig:Trajs}, so the geodesics producing UHECRs may be somewhat rare among the entire population (i.e., occupy a small volume in the space of geodesic initial data). They are however always possible. For example, \cite{2016CQGra..33q5001S} had shown through an analytical analysis of a simpler axisymmetric static surrogate model to the binary, that null geodesics (which our particles would approximately follow once they become highly relativistic) in a two-BH (dihole) system form a chaotic system, containing sequences of going-around-a-BH actions with arbitrary length (in fact, even infinitely lengthy ones where the geodesics do not ever escape or fall into either BH exist).    
The issue though, is how much rarer are the longer sequences. If the vanquish probability for a particle (due to it falling into the BHs or escaping the binary prematurely with too little energy) after each slingshot can be modelled as being a constant $\nu$, then trajectories fulfilling $N_{\rm min}^{0.01c}$ would be a factor $(1-\nu)^{900}$ ($\approx 10^{-271}$ if $\nu=1/2$) rarer than for those satisfying $N_{\rm min}^{0.1c}$, favouring the merger rather than the earlier inspiral phase as the stage of UHECR production, which implies a bursty, rather than steady, nature for UHECR sources. 

However, this constant $\nu$ assumption is not necessarily justified. Note first that for a particle to escape the binary, it is not sufficient to just have enough energy. The particle has to consistently move in the same outward direction over a period of time. Such escape attempts are however frustrated by the rapidly shifting gravitational potential in the vicinity of the binary, due to the BHs moving about. So the velocity of the particle becomes randomized (see Fig.~\ref{fig:Trajs}) and its escape more difficult. In fact, trapping is possible already under the less favourable conditions of a static dihole, for more quickly moving light rays following null geodesics, and the trapping orbits can even be stable \cite{2016PhRvD..94d4038D}. Also note that as compared to Fermi acceleration, trapping is much easier and more direct in the RS mechanism, since the scattering centres (BHs) themselves exert a long range attractive force. Instead, falling into the BHs likely represent a greater contribution to $\nu$ (this is what we see in our Newtonian simulation that produced Fig.~\ref{fig:Trajs}), and this contribution declines as a particle picks up speed, since it then becomes more light-like, thus more resilient to being pulled into a BH. In other words, $\nu$ likely declines as the number of slingshots already experienced by a particle builds up, causing further slingshots to become less difficult to achieve. Consequently, $\nu$ has a negative correlation with energy, which may explain the hardening of the UHECR spectrum as compared to galactic CRs, produced by first order Fermi acceleration with an energy-independent $\nu$ (see below Eq.~3 in \cite{1978MNRAS.182..147B}). If we instead concentrate on any particular energy and let binary separation vary, this consideration then ameliorates suppression due to large $N_{\rm min}$ numbers and opens up the possibility that BBHs in earlier inspiral phases end up winning out as the more proliferate sources, due to the greater population of BBH systems in this stage. We turn to this issue in the next section (before returning to some more discussion on the geodesics in Sec.~\ref{sec:Conclusion}).

\subsection{Astronomical binaries}
\subsubsection{Source population}\label{sec:EvRate}
From discussions in Sec.~\ref{sec:composition}, we recall that it is advantageous for those source binaries responsible for the harder ($\gtrsim 5$EeV) UHECRs to reside within the local universe ($\lesssim 1$Gpc), since then the photodisintegration of the C/N/O group elements provides a high energy cutoff in the UHECR flux density spectrum at precisely the correct place, without needing to invoke an extremely sharp intrinsic cutoff in the acceleration mechanism (due to its scarcity of length scales -- see Sec.~\ref{sec:Conclusion}, there is no new suppression mechanism turning on in the RS process that could cause such an abrupt cutoff at a specific energy). Indeed, the dipole in the galaxy distribution extracted from the 2-micron all-sky redshift survey extending to about $300$Mpc (see Fig.~3 in \cite{2006MNRAS.368.1515E}) is approximately consistent with the large scale UHECR anisotropy (see Fig.~3 in \cite{2017Sci...357.1266.}). 

Within such a moderately expansive volume, BBH systems are abundant. The aLIGO detector, without even reaching its design sensitivity ($\sim 1$Gpc for $M_{\rm tot}=100\text{M}_{\odot}$), and turned on for only a few months, had already seen five (and a sixth at lower confidence) BBH merger events (rate estimate is $2-600\text{Gpc}^{-3}\text{yr}^{-1}$, see \cite{2016ApJ...833L...1A}). The population of binaries still in their secular quasi-circular inspiral stage is thus quite substantial, since binaries spend much more time in this stage than the plunge-merger phase.
Specifically, a simple quadrupolar estimate on the rate of energy loss due to gravitational radiation, and subsequently orbital decay, yields for the time-until-merger
\bea \label{eq:TimeLeft}
T \propto \Omega^{-8/3} \,,
\eea
where the proportionality coefficient is such that two $30M_{\odot}$ BHs will have around one minute till merger when they enter the LIGO and VIRGO band at a GW frequency of around $f=2\times \Omega/(2\pi)\approx 10$Hz (translating to around $b=20M_{\rm tot}$ by Eq.~\eqref{eq:Kepler}). 
In a steady state population distribution then, the number density $\rho$ per unit frequency range at any given $f$ must satisfy (by continuity equation with a steady injection from binary formation, and a steady removal due to mergers)
\bea \label{eq:nScaling}
\rho\propto f^{-11/3}\,.
\eea 
Taking the geometric mean (averaging over orders of magnitude) of the most optimistic and pessimistic event (having a BBH in band) rate estimates, at (still quite conservative given the number of events already seen) $\sim 35\text{Gpc}^{-3}\text{yr}^{-1}$ (i.e., total population density of BBHs with less than one year to merger, or $b \lesssim 550M_{\rm tot}$ for $M_{\rm tot}=60M_{\odot}$, is $35\text{Gpc}^{-3}$), we obtain that the proportionality coefficient in Eq.~\eqref{eq:nScaling} should give
\begin{align}
\rho\left(f=1\text{Hz}\right)
=\frac{1}{8}\text{Gpc}^{-3}\text{Hz}^{-1}\,,
\end{align}
at any moment in time. Using Eqs.~\eqref{eq:nScaling}, \eqref{eq:TimeLeft} and \eqref{eq:Kepler}, we can then compute the relevant numbers corresponding to various stages of the BBH coalescence, which are displayed in Tb.~1 (note that $b\approx 3000M_{\rm tot}$ is less than half of the solar radius, so the binary orbit is still quite tight even in this case). 

\begin{table}
  \centering \label{tb:params}
  \begin{tabular}{ccccc}
    \hline \hline
    $b/M_{\rm tot}$ & $f/{\rm Hz}$ & $|{\bf v_A}|/c$ & $T$ & $P^{\rm op}/\text{Gpc}^{-3}$\\
    \hline
    3000 & 0.007 & 0.009 & $963$yr & $3\times 10^4$ \\
    \midrule    
    2000 & 0.01 & 0.01 & $190$yr & $6\times 10^3$ \\
    \midrule
    550 & 0.08 & 0.02 & $1$yr & $35$ \\
    \midrule
    200 & 0.4 & 0.04 & $7$d & $0.6$ \\
    \midrule
    20 & 12 & 0.1 & $1$min & $6 \times 10^{-5}$ \\
    \hline \hline
  \end{tabular}
  \caption{Approximate GW frequency $f$, BH speed $|{\bf v_A}|$, and time-until-merger $T$ corresponding to various binary separations $b$, as well as the population $P^{\rm op}$ of all binaries with separation equal to and below that $b$ value heading the same row. }
\end{table}

For $b \lesssim 550M_{\rm tot}$, we essentially have a transient burst source, corresponding to the BBH merger events, happening at a low frequency of $2-600\text{Gpc}^{-3}\text{yr}^{-1}$. This demands a certain luminosity from each source, in order to account for the rate of UHECR observations on Earth. More quantitatively, an approximation for the flux seen on Earth at above $10^{19}$eV is given in \cite{2011ARA&A..49..119K} at 
\begin{align} \label{eq:BurstScaling}
\left(E^3\frac{dN}{dE}\right)\Bigg|_{E=10^{19}\text{eV}} \sim 
&10^{24} \text{eV}^2\text{m}^{-2}\text{s}^{-1}\text{sr}^{-1}
\left(\frac{\dot{P}^{\rm op}}{1 \text{Gpc}^{-3}\text{yr}^{-1}}\right) 
\left(\frac{E_{\geq 19}}{3\times 10^{53}\text{erg}}\right)\,, 
\end{align}
where $\dot{P}^{\rm op}$ is the burst event rate and $E_{\geq 19}$ is the total energy injected into UHECR at above $10^{19}$eV. The second line of Eq.~\eqref{eq:BurstScaling} should be approximately unity to match observations (see Fig.~2 of \cite{2011ARA&A..49..119K}). The total energy of $\sim 10^{51}-10^{53}$erg is quite formidable, but still smaller than the $\sim 5\times 10^{54}$erg injected into GW  (both the slingshots and GW ultimately get their energy from the dynamical gravitational field surrounding the BHs) during the merger event \cite{2016PhRvL.116x1103A}, so the overall energy budget is in principle available. The more pressing issue is whether there exists a large enough population of carrier nuclei to propagate this energy out. Fortunately, the energy per particle is extremely high, at $\sim 10^{19}$eV in accordance with Eq.~\eqref{eq:BurstScaling}, so the total number of particles is only roughly $5\times 10^{44}$ per burst event, translating to around $10^{-6}-10^{-5}$ times the mass of the Moon (or $10^{-13}-10^{-12}M_{\odot}$) depending on whether the particles are protons or C/N/O nuclei. An accretion disk (likely much more tenuous than those seen in e.g., X-ray binaries, and without the dense disk processes like thermal X-ray emissions) forming out of fallback material of the hypernovae explosions \cite{2014ApJ...781..119P} or tidally disrupted objects \cite{2011MNRAS.415.3824S} (in particular, in the popular dynamical evolution channel for the formation of extremely tight BBH systems that can merge within the age of the universe, interloper stars are required to thread through the -- already rather close -- binary to take away orbital energy, and can become partial tidally disrupted in the process, see \cite{2018arXiv180605820M} for a concise introduction and references), and kept around (prevented from being quickly accreted by the BHs) by the binary tidal torque and/or the existence of a dead zone \cite{DiskDeadZone}, could thus plausibly supply the required particle number. 

For the accretion scenario to operate effectively though, one has to consider the constraints laid down by Eddington luminosity. However, it is not the UHECR luminosity that's the relevant one, because the UHECR energy is concentrated on a very small number of particles, as opposed to a large number of low energy photons, so there is little chance for the infalling material to collide with them. In other words, the UHECR particles streaming outwards will not produce a pressure that shuts down accretion. The relevant luminosities are instead those associated with the lower-energy-per-particle radiation, such as X-ray or $\gamma$-ray, carried by swarms of particles (e.g., photons) that permeate the BBH region. Studies have shown that these luminosities (they are not produced by the RS mechanism, and are independent of the UHECR luminosity) are quite low (see e.g., \cite{2017MNRAS.465.4406K,2017ApJ...839L...7D} for theoretical studies, and also note that the lack of EM counterparts to BBH mergers places direct observational constraints). 
If this bursty scenario is dominant though, one could potentially hope for a correlation between UHECR and GW observations, with one signature of the RS mechanism being that the UHECRs temporally precede the GW (but deflection of the CR particles by magnetic fields may cause them to take a detour, so careful data analysis is needed). So far, efforts had been made to search for photons and neutrinos coincident with GW from a double neutron star merger \cite{2017ApJ...848L..12A,2017ApJ...850L..35A}, and if BBHs are indeed the sources of UHECRs, two new fronts in multimessenger astronomy may open up, namely the utilization of charged particles as messengers, as well as with BBHs as observation targets.

Alternatively, as discussed at the end of Sec.~\ref{sec:Params}, it may not necessarily be prohibitively more difficult to find geodesics making many more slingshots. If this is true, then the inspiral phase of the BBH, lasting hundreds of years even as the binary is already very tight, would serve as a steady/continuous source to the UHECRs. Moreover, the number density of such BBH systems is very high (first two rows of Tb.~1), which contrasts with other possible steady sources such as clusters of galaxies or radio galaxies (at $P^{\rm op}\sim 1 - 10\text{Gpc}^{-3}$, see e.g., \cite{2005A&A...434..133W}), and helps the RS mechanism to better reconcile with the lack of significant small scale clustering in observed UHECR directions \cite{Abreu:2013kif,2015ApJ...804...15A}. Incidentally, the population density for $b \lesssim 3000 M_{\rm tot}$, as displayed in the first row of Tb.~1, is just above the lower bound set by \cite{Abreu:2013kif} according to clustering considerations (when assuming magnetic deflection angles appropriate for a mixed composition). Thanks to this large population, the demand on particle supply at each site is also lower in the steady case, for which \cite{2011ARA&A..49..119K} estimates
\begin{align}
\left(E^3\frac{dN}{dE}\right)\Bigg|_{E=10^{19}\text{eV}} \sim 
& 10^{24} \text{eV}^2\text{m}^{-2}\text{s}^{-1}\text{sr}^{-1}
\left(\frac{P^{\rm op}}{10^{4} \text{Gpc}^{-3}}\right) 
\left(\frac{L_{19}}{10^{42}\text{erg\, s}^{-1}}\right)\,, 
\end{align}
where $L_{19}$ is the steady luminosity above $10^{19}$eV from each source system. Substituting in the population estimate for $b\lesssim 3000M_{\rm tot}$, we have that $\sim 2\times 10^{34}$ nuclei (equalling the rest mass of $25-50\text{m}^3$ of Earth rock) must escape each BBH per second. One may wonder if this opens up the possibility that even without accretion disks, interstellar medium may suffice. We do not deem this scenario plausible, since at the average one particle per $\text{cm}^3$ interstellar medium density, the flux would clean out the BBH region $10\times 3000 M_{\rm tot}$ (slightly larger than the size of the sun) across in just a few seconds. Although with a density of $10^{6}\text{cm}^{-3}$ in cool dense interstellar medium regions, this time scale grows to about a month, we have to take caution that the estimate is done without accounting for the particles falling into the holes. Therefore, accretion is still likely required. 
If this steady scenario is dominant, then we would not expect coincident observations by UHECR and ground based GW detectors, since the BBH systems would only enter into the sensitive bands of the latter at the very last stages of inspiral, just a few cycles before merger. In principal, space-based gravitational wave detectors with appropriate arm-lengths may be able to see the inspiral phase, but given the very large number of sources, they likely will only see them as a stochastic background and not disentangled individual sources, so this avenue for establishing GW counterparts to UHECR may not be easy, unless we are lucky that one or few sources are particularly close by.

\subsubsection{Competing energy draining processes}
Besides the main accelerating RS mechanism that's the focus of this paper, we must caution that just as with any other proposals on the UHECR sources, there could be a myriad of other processes also at work in the source region, which could drain energy out of the particles, resulting in more slingshots being required to reach the UHECR energies.

The most familiar of such energy loss channels is perhaps synchrotron radiation, which for a nuclei gives an energy loss time scale inversely proportional to the magnetic field strength squared \cite{1984ARA&A..22..425H} 
\bea \label{eq:Synch}
t_S = \left(\frac{A}{Z}\right)^4 \frac{1.4}{E_{20}B^2} \text{yr}\,,
\eea
where $B$ is in units of Gauss and $E_{20}$ in multiples of $10^{20}$eV. This tends to be a serious issue with EM acceleration mechanisms (e.g., it represents an additional constraint on the Hillas-plot), since the accelerating process itself demands a strong magnetic field. The RS or other gravity-based processes thus possess an advantage in that $B$ can be tuned down without affecting the energization mechanism. With Eq.~\eqref{eq:Synch}, we can estimate an upper bound for $B$ below which synchrotron losses will not become a significant factor during the RS process. For the steady source case, the time needed to achieve $N^{0.01c}_{\rm min}$ light crossings across $b=3000M_{\rm tot}$ is around half an hour, and for $t_S$ to be larger than this (i.e., for the loss time to be greater than acceleration time) we would need $B<40$G. For the bursty sources, the constrain is even looser, the light crossing time for $N^{0.1c}_{\rm min}$ and $b=20M_{\rm tot}$ is $1.2$s, leading to $B<1500$G. With a BBH system, there is little observational clue regarding the strength of a surrounding magnetic field (possibly supported by currents within a circumbinary accretion disk).
Nevertheless, we can take some cues from the situation where a single BH accretes from a companion star. When a particularly strong burst of accretion happened for V404 Cygni, the magnetic field strength in the synchrotron radiation region has been measured to be $33$G \cite{2017Sci...358.1299D}, which is comparable to the steady case limit computed with the idealized $N^{0.01c}_{\rm min}$. If such a field strength is indeed comparable to that around a BBH then (also note that magneto-rotational instability could possibly further boost the field strength beyond that of a steady accretion), it could be argued that bursty UHECR sources are more favourable.

The other commonly evoked energy loss channel in the UHECR context is the inverse-Compton scattering against photons. This loss can not be estimated without ascertaining the photon background. But we can once again estimate upper limits above which the energy loss time scale due to this effect becomes smaller than the acceleration time scale. 
The ratio between the energy loss rates for inverse-Compton scattering and synchrotron radiation is normally approximately equal to photon energy density $u_{\rm rad}$ over magnetic field energy density $u_{\rm mag}$. So we can take the results from the last paragraph as starting points, and use the black body expression  
\bea
u_{\rm rad} = \frac{8\pi^5(k_BT)^4}{15(hc)^3}\,, 
\eea
to obtain the threshold radiation temperatures of $\sim 10^5$K for the steady sources and $\sim 10^6$K for the bursty sources. The temperature of the Sun is around $6\times 10^4$K (photosphere) to $2\times 10^7$K (centre), so taken at face value, our estimates suggest that a freshly disrupted hot star is less favourable as a supply for UHECR candidate particles than cold accretion disks forming (a long time ago) out of hypernovae fall-back material. One should be careful however, that the naive energetic argument above only leads to a very generous upper limit for how important inverse-Compton scattering can become. In reality, at the extremely high charged particle energies (in its rest frame, the photon is strongly blue shifted) relevant for UHECRs, the Klein-Nishina effect would strongly suppress the scattering cross section.

\section{Conclusion} \label{sec:Conclusion}
In this paper, we derived general relativistic slingshot formulae with the aim to preliminarily investigate the possibility that repeated slingshots around the BHs in a tight binary may be able to accelerate accreting particles into UHECRs. Our computations show that there is no degradation to the effectiveness of the slingshot mechanism even as the particles already possess extreme energies. Therefore, the RS mechanism overcomes perhaps the most rudimentary and difficult issue facing any UHECR source theory, namely that the accelerating mechanism may shut down at a threshold energy below the incredible macroscopic values recorded observationally. For example, if acceleration is accomplished with an electrostatic field instead, then once the energy of the particles rises above a certain level then, the photon energy will become greater than the rest mass of an electron-positron pair (only $1$MeV, much less than the energies of the UHECR), causing a cascade of pair production that shorts out the electric field and thus shuts down further acceleration. On the other hand, the ability of the RS mechanism to avoid such an energy ceiling is not entirely surprising, given the particularly simple physics involved. One notes that in geometrized units, energy carries the dimension of length, but there is really only one length scale $M_{\rm tot}$ in the equal-mass binary problem (this is why one does not need to simulate BBH systems differing only in $M_{\rm tot}$ when making GW templates, since one can simply rescale all results according to their dimensions), so there really isn't many ways to construct such an energy ceiling. In contrast, in the EM alternative, one would have the charge and mass of the particles, the electric and magnetic field strengths at the acceleration site, the radius of curvature of the magnetic field etc all available for one to make up many quantities with the dimension of energy, which can then plausibly serve as the energy ceiling. 

Unfortunately however, it is not possible to analytically string up the individual slingshots into complete chaotic geodesics, and examine how the population of particles depends on the total number of slingshots experienced (the jaggedness of numerical data in Figs.~1 and 2 of Ref.~\cite{2010ApJ...711L..89V} hints at how difficult this task would be). As a result, we haven't been able to produce a detailed source energy spectrum in this paper, and can not ascertain whether the UHECR sources should be bursty or steady within the RS framework. 
Indeed, given these limitations in our analysis, we can not claim that the RS mechanism is viable as an explanation for a \emph{dominant proportion} of the UHECRs produced in the universe. It may well turn out to only be of cursory importance. However, we think RS is sufficiently interesting in its own right as an acceleration mechanism, that it is worthwhile presenting and analysing it to the fullest of our abilities, and hopefully garner interest from more capable investigators. To make more realistic assessments on just how much of the observable CRs observed on earth, if any, were produced by RS, much more sophisticated investigations are required, which are perhaps best performed with fully nonlinear numerical BBH simulations, because we need to examine the late inspiral and plunge-merge phases, for which post-Newtonian expressions may not be sufficient, and also particularly because the relevant geodesics will sample the regions that are the strongest in gravity and also the most dynamic. The true BBH system being dynamic also means that the analytic technique of approximating the binary by the static and axisymmetric Majumdar-Parapetrou spacetime, which produced some of the qualitative features that we quoted in Sec.~\ref{sec:Params}, is no longer applicable for a quantitative study. On the other hand, large separation $b$ values at hundreds or thousands of $M_{\rm tot}$ may be quite challenging to simulate numerically, since the gravitational attraction between the two BHs is weak and numerical errors, especially those arising from the initial junk radiation (the initial data of the simulation are not precise, so the spacetime needs to settle down into the physical solution by first radiating away the errors of the man-made initial data; this radiation can contain high frequency components that are under-resolved, see e.g., \cite{2013PhRvD..88h4033Z}) can possibly send the two BHs flying apart. So a post-Newtonian NZ (see Fig.~\ref{fig:Zones}) glued onto two boosted Kerr-Schild (BHs are not singular on the horizon in this coordinate system, thus more appropriate for numerics) IZs may be the best that's currently practical for this case. 

Beyond the binary metric, one would also need to launch a vast number of geodesics into the simulated spacetime (perhaps similar to \cite{2015CQGra..32f5002B} and \cite{2010ApJ...711L..89V}, but with many more of them), in order to carry out a population study to find out how much accretion onto the binary is needed to provide sufficient luminosity, and acquire a detailed prediction for the spectrum. This could be done post-processing, but the coding has to be very careful to avoid all these geodesics separately and repeatedly carry out time-consuming input/output operations to read in stored metric data from the hard drives. A better strategy may be to integrate the geodesics concurrently with the main spacetime simulation, but this may require alterations to the various fundamental utilities in the code such as those handling time-stepping. Nevertheless, we expect rich rewards for experts willing to carry out this task. 
Due to the BH motions being regular, the phase space of geodesics in a binary likely possess clean organizational structures (i.e., the distribution of geodesic properties won't be completely random, so we have refrained from making oversimplifying statistical assumptions such as ``complete randomization in the impact parameter between consecutive slingshots'' that would undoubtedly result in wrong predictions for the UHECR spectrum), perhaps similar to those found by \cite{2016CQGra..33q5001S} and studies of geodesics in the even simpler but also axisymmetric Kerr spacetime (see e.g., \cite{2015PhRvD..91h3001B}), but which can only be richer in structure with a truly dynamical BBH system given the reduced symmetry, and thus reduced number of conserved quantities constraining the geodesics.
In other words, it is likely that the geodesics capable of producing UHECRs correspond to particular classes of dynamical systems concepts, in which case it may become possible to put the studies on spectrum and emission skymap etc on a quite solid mathematical footing, generating precise testable predictions. We hope our proposed astrophysical relevance would help give some impetus to this study.

\vspace{10mm}
This work is supported by the National Natural Science Foundation of China grants 11503003 and 11633001, the Interdiscipline Research Funds of Beijing Normal University, the Strategic Priority Research Program of the Chinese Academy of Sciences Grant No. XDB23000000, the Fundamental Research Funds for the Central Universities grant 2015KJJCB06, and a Returned Overseas Chinese Scholars Foundation grant. 

\bibliographystyle{spphys}       
\bibliography{slingshot.bbl}   

\begin{thebibliography}{10}
\providecommand{\url}[1]{{#1}}
\providecommand{\urlprefix}{URL }
\expandafter\ifx\csname urlstyle\endcsname\relax
  \providecommand{\doi}[1]{DOI \discretionary{}{}{}#1}\else
  \providecommand{\doi}{DOI \discretionary{}{}{}\begingroup
  \urlstyle{rm}\Url}\fi

\bibitem{1961PhRvL...6..485L}
J.~{Linsley}, L.~{Scarsi}, B.~{Rossi}, Physical Review Letters \textbf{6}, 485
  (1961).
\newblock \doi{10.1103/PhysRevLett.6.485}

\bibitem{1963PhRvL..10..146L}
J.~{Linsley}, Physical Review Letters \textbf{10}, 146 (1963).
\newblock \doi{10.1103/PhysRevLett.10.146}

\bibitem{2012APh....35..660K}
K.H. {Kampert}, M.~{Unger}, Astroparticle Physics \textbf{35}, 660 (2012).
\newblock \doi{10.1016/j.astropartphys.2012.02.004}

\bibitem{2017AIPC.1852d0001K}
K.H. {Kampert}, in \emph{American Institute of Physics Conference Series},
  \emph{American Institute of Physics Conference Series}, vol. 1852 (2017),
  \emph{American Institute of Physics Conference Series}, vol. 1852, p. 040001.
\newblock \doi{10.1063/1.4984858}

\bibitem{2014PhRvD..90l2005A}
A.~{Aab}, P.~{Abreu}, M.~{Aglietta}, E.J. {Ahn}, I.~{Al Samarai}, I.F.M.
  {Albuquerque}, I.~{Allekotte}, J.~{Allen}, P.~{Allison}, A.~{Almela}, et~al.,
  Phys. Rev. D \textbf{90}(12), 122005 (2014).
\newblock \doi{10.1103/PhysRevD.90.122005}

\bibitem{2014PhRvD..90l2006A}
A.~{Aab}, P.~{Abreu}, M.~{Aglietta}, E.J. {Ahn}, I.~{Al Samarai}, I.F.M.
  {Albuquerque}, I.~{Allekotte}, J.~{Allen}, P.~{Allison}, A.~{Almela}, et~al.,
  Phys. Rev. D \textbf{90}(12), 122006 (2014).
\newblock \doi{10.1103/PhysRevD.90.122006}

\bibitem{2015APh....64...49A}
R.U. {Abbasi}, M.~{Abe}, T.~{Abu-Zayyad}, M.~{Allen}, R.~{Anderson},
  R.~{Azuma}, E.~{Barcikowski}, J.W. {Belz}, D.R. {Bergman}, S.A. {Blake},
  R.~{Cady}, M.J. {Chae}, B.G. {Cheon}, J.~{Chiba}, M.~{Chikawa}, W.R. {Cho},
  T.~{Fujii}, M.~{Fukushima}, T.~{Goto}, W.~{Hanlon}, Y.~{Hayashi},
  N.~{Hayashida}, K.~{Hibino}, K.~{Honda}, D.~{Ikeda}, N.~{Inoue}, T.~{Ishii},
  R.~{Ishimori}, H.~{Ito}, D.~{Ivanov}, C.C.H. {Jui}, K.~{Kadota},
  F.~{Kakimoto}, O.~{Kalashev}, K.~{Kasahara}, H.~{Kawai}, S.~{Kawakami},
  S.~{Kawana}, K.~{Kawata}, E.~{Kido}, H.B. {Kim}, J.H. {Kim}, J.H. {Kim},
  S.~{Kitamura}, Y.~{Kitamura}, V.~{Kuzmin}, Y.J. {Kwon}, J.~{Lan}, S.I. {Lim},
  J.P. {Lundquist}, K.~{Machida}, K.~{Martens}, T.~{Matsuda}, T.~{Matsuyama},
  J.N. {Matthews}, M.~{Minamino}, Y.~{Mukai}, I.~{Myers}, K.~{Nagasawa},
  S.~{Nagataki}, T.~{Nakamura}, T.~{Nonaka}, A.~{Nozato}, S.~{Ogio},
  J.~{Ogura}, M.~{Ohnishi}, H.~{Ohoka}, K.~{Oki}, T.~{Okuda}, M.~{Ono},
  A.~{Oshima}, S.~{Ozawa}, I.H. {Park}, M.S. {Pshirkov}, D.C. {Rodriguez},
  G.~{Rubtsov}, D.~{Ryu}, H.~{Sagawa}, N.~{Sakurai}, A.L. {Sampson}, L.M.
  {Scott}, P.D. {Shah}, F.~{Shibata}, T.~{Shibata}, H.~{Shimodaira}, B.K.
  {Shin}, H.S. {Shin}, J.D. {Smith}, P.~{Sokolsky}, R.W. {Springer}, B.T.
  {Stokes}, S.R. {Stratton}, T.~{Stroman}, T.~{Suzawa}, M.~{Takamura},
  M.~{Takeda}, R.~{Takeishi}, A.~{Taketa}, M.~{Takita}, Y.~{Tameda},
  H.~{Tanaka}, K.~{Tanaka}, M.~{Tanaka}, S.B. {Thomas}, G.B. {Thomson},
  P.~{Tinyakov}, I.~{Tkachev}, H.~{Tokuno}, T.~{Tomida}, S.~{Troitsky},
  Y.~{Tsunesada}, K.~{Tsutsumi}, Y.~{Uchihori}, S.~{Udo}, F.~{Urban},
  G.~{Vasiloff}, T.~{Wong}, R.~{Yamane}, H.~{Yamaoka}, K.~{Yamazaki},
  J.~{Yang}, K.~{Yashiro}, Y.~{Yoneda}, S.~{Yoshida}, H.~{Yoshii},
  R.~{Zollinger}, Z.~{Zundel}, Astroparticle Physics \textbf{64}, 49 (2015).
\newblock \doi{10.1016/j.astropartphys.2014.11.004}

\bibitem{2012ApJ...749...63M}
K.~{Murase}, C.D. {Dermer}, H.~{Takami}, G.~{Migliori}, Astrophys. J.
  \textbf{749}, 63 (2012).
\newblock \doi{10.1088/0004-637X/749/1/63}

\bibitem{2010PhRvL.104i1101A}
J.~{Abraham}, P.~{Abreu}, M.~{Aglietta}, E.J. {Ahn}, D.~{Allard},
  I.~{Allekotte}, J.~{Allen}, J.~{Alvarez-Mu{\~n}iz}, M.~{Ambrosio},
  L.~{Anchordoqui}, et~al., Physical Review Letters \textbf{104}(9), 091101
  (2010).
\newblock \doi{10.1103/PhysRevLett.104.091101}

\bibitem{2018ApJ...853L..29A}
A.~{Aab}, P.~{Abreu}, M.~{Aglietta}, I.F.M. {Albuquerque}, I.~{Allekotte},
  A.~{Almela}, J.~{Alvarez Castillo}, J.~{Alvarez-Mu{\~n}iz}, G.A. {Anastasi},
  L.~{Anchordoqui}, et~al., Astrophys. J. Lett. \textbf{853}, L29 (2018).
\newblock \doi{10.3847/2041-8213/aaa66d}

\bibitem{2018arXiv181101108F}
N.~{Fraija}, M.~{Araya}, A.~{Galvan-Gamez}, J.A. {de Diego}, arXiv e-prints
  (2018)

\bibitem{2018Sci...361.1378I}
{IceCube Collaboration}, M.G. {Aartsen}, M.~{Ackermann}, J.~{Adams}, J.A.
  {Aguilar}, M.~{Ahlers}, M.~{Ahrens}, I.~{Al Samarai}, D.~{Altmann},
  K.~{Andeen}, et~al., Science \textbf{361}, eaat1378 (2018).
\newblock \doi{10.1126/science.aat1378}

\bibitem{2018Sci...361..147I}
{IceCube Collaboration}, M.G. {Aartsen}, M.~{Ackermann}, J.~{Adams}, J.A.
  {Aguilar}, M.~{Ahlers}, M.~{Ahrens}, I.A. {Samarai}, D.~{Altmann},
  K.~{Andeen}, et~al., Science \textbf{361}, 147 (2018).
\newblock \doi{10.1126/science.aat2890}

\bibitem{2014PhRvD..90b3007M}
K.~{Murase}, Y.~{Inoue}, C.D. {Dermer}, Phys. Rev. D \textbf{90}(2), 023007
  (2014).
\newblock \doi{10.1103/PhysRevD.90.023007}

\bibitem{2018arXiv180704537K}
A.~{Keivani}, K.~{Murase}, M.~{Petropoulou}, D.B. {Fox}, S.B. {Cenko},
  S.~{Chaty}, A.~{Coleiro}, J.J. {DeLaunay}, S.~{Dimitrakoudis}, P.A. {Evans},
  J.A. {Kennea}, F.E. {Marshall}, A.~{Mastichiadis}, J.P. {Osborne},
  M.~{Santander}, A.~{Tohuvavohu}, C.F. {Turley}, ArXiv e-prints  (2018)

\bibitem{2018arXiv180704275G}
S.~{Gao}, A.~{Fedynitch}, W.~{Winter}, M.~{Pohl}, ArXiv e-prints  (2018)

\bibitem{1995PhRvL..75..386W}
E.~{Waxman}, Physical Review Letters \textbf{75}, 386 (1995).
\newblock \doi{10.1103/PhysRevLett.75.386}

\bibitem{1995ApJ...449L..37M}
M.~{Milgrom}, V.~{Usov}, Astrophys. J. Lett. \textbf{449}, L37 (1995).
\newblock \doi{10.1086/309633}

\bibitem{2012Natur.484..351A}
R.~{Abbasi}, Y.~{Abdou}, T.~{Abu-Zayyad}, M.~{Ackermann}, J.~{Adams}, J.A.
  {Aguilar}, M.~{Ahlers}, D.~{Altmann}, K.~{Andeen}, J.~{Auffenberg}, et~al.,
  Nature \textbf{484}, 351 (2012).
\newblock \doi{10.1038/nature11068}

\bibitem{2016PhRvL.116f1102A}
B.P. {Abbott}, R.~{Abbott}, T.D. {Abbott}, M.R. {Abernathy}, F.~{Acernese},
  K.~{Ackley}, C.~{Adams}, T.~{Adams}, P.~{Addesso}, R.X. {Adhikari}, et~al.,
  Physical Review Letters \textbf{116}(6), 061102 (2016).
\newblock \doi{10.1103/PhysRevLett.116.061102}

\bibitem{2016PhRvL.116x1103A}
B.P. {Abbott}, R.~{Abbott}, T.D. {Abbott}, M.R. {Abernathy}, F.~{Acernese},
  K.~{Ackley}, C.~{Adams}, T.~{Adams}, P.~{Addesso}, R.X. {Adhikari}, et~al.,
  Physical Review Letters \textbf{116}(24), 241103 (2016).
\newblock \doi{10.1103/PhysRevLett.116.241103}

\bibitem{2016ApJ...833L...1A}
B.P. {Abbott}, R.~{Abbott}, T.D. {Abbott}, M.R. {Abernathy}, F.~{Acernese},
  K.~{Ackley}, C.~{Adams}, T.~{Adams}, P.~{Addesso}, R.X. {Adhikari}, et~al.,
  Astrophys. J. Lett. \textbf{833}, L1 (2016).
\newblock \doi{10.3847/2041-8205/833/1/L1}

\bibitem{2017PhRvL.118v1101A}
B.P. {Abbott}, R.~{Abbott}, T.D. {Abbott}, F.~{Acernese}, K.~{Ackley},
  C.~{Adams}, T.~{Adams}, P.~{Addesso}, R.X. {Adhikari}, V.B. {Adya}, et~al.,
  Physical Review Letters \textbf{118}(22), 221101 (2017).
\newblock \doi{10.1103/PhysRevLett.118.221101}

\bibitem{2017ApJ...851L..35A}
B.P. {Abbott}, R.~{Abbott}, T.D. {Abbott}, F.~{Acernese}, K.~{Ackley},
  C.~{Adams}, T.~{Adams}, P.~{Addesso}, R.X. {Adhikari}, V.B. {Adya}, et~al.,
  Astrophys. J. Lett. \textbf{851}, L35 (2017).
\newblock \doi{10.3847/2041-8213/aa9f0c}

\bibitem{2017PhRvL.119n1101A}
B.P. {Abbott}, R.~{Abbott}, T.D. {Abbott}, F.~{Acernese}, K.~{Ackley},
  C.~{Adams}, T.~{Adams}, P.~{Addesso}, R.X. {Adhikari}, V.B. {Adya}, et~al.,
  Physical Review Letters \textbf{119}(14), 141101 (2017).
\newblock \doi{10.1103/PhysRevLett.119.141101}

\bibitem{2010ApJ...711L..89V}
J.R. {van Meter}, J.H. {Wise}, M.C. {Miller}, C.S. {Reynolds}, J.~{Centrella},
  J.G. {Baker}, W.D. {Boggs}, B.J. {Kelly}, S.T. {McWilliams}, Astrophys. J.
  Lett. \textbf{711}, L89 (2010).
\newblock \doi{10.1088/2041-8205/711/2/L89}

\bibitem{1938RSPSA.167..148D}
P.A.M. {Dirac}, Proceedings of the Royal Society of London Series A
  \textbf{167}, 148 (1938).
\newblock \doi{10.1098/rspa.1938.0124}

\bibitem{1999gr.qc....12045P}
E.~{Poisson}, ArXiv General Relativity and Quantum Cosmology e-prints  (1999)

\bibitem{1997PhRvD..56.3381Q}
T.C. {Quinn}, R.M. {Wald}, Phys. Rev. D \textbf{56}, 3381 (1997).
\newblock \doi{10.1103/PhysRevD.56.3381}

\bibitem{1949PhRv...75.1169F}
E.~{Fermi}, Physical Review \textbf{75}, 1169 (1949).
\newblock \doi{10.1103/PhysRev.75.1169}

\bibitem{1977ICRC...11..132A}
W.I. {Axford}, E.~{Leer}, G.~{Skadron}, International Cosmic Ray Conference
  \textbf{11}, 132 (1977)

\bibitem{1978MNRAS.182..147B}
A.R. {Bell}, Mon. Not. R. Astron. Soc. \textbf{182}, 147 (1978).
\newblock \doi{10.1093/mnras/182.2.147}

\bibitem{1978ApJ...221L..29B}
R.D. {Blandford}, J.P. {Ostriker}, Astrophys. J. Lett. \textbf{221}, L29
  (1978).
\newblock \doi{10.1086/182658}

\bibitem{1955RSPSA.229..416B}
H.~{Bondi}, T.~{Gold}, Proceedings of the Royal Society of London Series A
  \textbf{229}, 416 (1955).
\newblock \doi{10.1098/rspa.1955.0098}

\bibitem{1963AnPhy..22..169R}
F.~{Rohrlich}, Annals of Physics \textbf{22}, 169 (1963).
\newblock \doi{10.1016/0003-4916(63)90051-4}

\bibitem{Gron:2012pr}
O.~Gron, Adv. Math. Phys. \textbf{2012}, 528631 (2012).
\newblock \doi{10.1155/2012/528631}

\bibitem{2017Sci...357.1266.}
{The Pierre Auger Collaboration}, Science \textbf{357}, 1266 (2017).
\newblock \doi{10.1126/science.aan4338}

\bibitem{2013Sci...342..334S}
R.M. {Shannon}, V.~{Ravi}, W.A. {Coles}, G.~{Hobbs}, M.J. {Keith}, R.N.
  {Manchester}, J.S.B. {Wyithe}, M.~{Bailes}, N.D.R. {Bhat},
  S.~{Burke-Spolaor}, J.~{Khoo}, Y.~{Levin}, S.~{Oslowski}, J.M. {Sarkissian},
  W.~{van Straten}, J.P.W. {Verbiest}, J.B. {Wang}, Science \textbf{342}, 334
  (2013)

\bibitem{2016ApJ...821...13A}
Z.~{Arzoumanian}, A.~{Brazier}, S.~{Burke-Spolaor}, S.J. {Chamberlin},
  S.~{Chatterjee}, B.~{Christy}, J.M. {Cordes}, N.J. {Cornish}, K.~{Crowter},
  P.B. {Demorest}, X.~{Deng}, T.~{Dolch}, J.A. {Ellis}, R.D. {Ferdman},
  E.~{Fonseca}, N.~{Garver-Daniels}, M.E. {Gonzalez}, F.~{Jenet}, G.~{Jones},
  M.L. {Jones}, V.M. {Kaspi}, M.~{Koop}, M.T. {Lam}, T.J.W. {Lazio},
  L.~{Levin}, A.N. {Lommen}, D.R. {Lorimer}, J.~{Luo}, R.S. {Lynch}, D.R.
  {Madison}, M.A. {McLaughlin}, S.T. {McWilliams}, C.M.F. {Mingarelli}, D.J.
  {Nice}, N.~{Palliyaguru}, T.T. {Pennucci}, S.M. {Ransom}, L.~{Sampson}, S.A.
  {Sanidas}, A.~{Sesana}, X.~{Siemens}, J.~{Simon}, I.H. {Stairs}, D.R.
  {Stinebring}, K.~{Stovall}, J.~{Swiggum}, S.R. {Taylor}, M.~{Vallisneri},
  R.~{van Haasteren}, Y.~{Wang}, W.W. {Zhu}, {NANOGrav Collaboration},
  Astrophys. J. \textbf{821}, 13 (2016).
\newblock \doi{10.3847/0004-637X/821/1/13}

\bibitem{2015MNRAS.453.2576L}
L.~{Lentati}, S.R. {Taylor}, C.M.F. {Mingarelli}, A.~{Sesana}, S.A. {Sanidas},
  A.~{Vecchio}, R.N. {Caballero}, K.J. {Lee}, R.~{van Haasteren}, S.~{Babak},
  C.G. {Bassa}, P.~{Brem}, M.~{Burgay}, D.J. {Champion}, I.~{Cognard},
  G.~{Desvignes}, J.R. {Gair}, L.~{Guillemot}, J.W.T. {Hessels}, G.H.
  {Janssen}, R.~{Karuppusamy}, M.~{Kramer}, A.~{Lassus}, P.~{Lazarus},
  K.~{Liu}, S.~{Os{\l}owski}, D.~{Perrodin}, A.~{Petiteau}, A.~{Possenti}, M.B.
  {Purver}, P.A. {Rosado}, R.~{Smits}, B.~{Stappers}, G.~{Theureau},
  C.~{Tiburzi}, J.P.W. {Verbiest}, Mon. Not. R. Astron. Soc. \textbf{453}, 2576
  (2015).
\newblock \doi{10.1093/mnras/stv1538}

\bibitem{2018arXiv180106160T}
{The Pierre Auger Collaboration}, A.~{Aab}, P.~{Abreu}, M.~{Aglietta}, I.F.M.
  {Albuquerque}, I.~{Allekotte}, A.~{Almela}, J.~{Alvarez Castillo},
  J.~{Alvarez-Mu{\~n}iz}, G.A. {Anastasi}, et~al., ArXiv e-prints  (2018)

\bibitem{2017ApJ...836...50L}
S.~{Laycock}, R.~{Cappallo}, B.F. {Williams}, A.~{Prestwich}, B.~{Binder}, D.M.
  {Christodoulou}, Astrophys. J. \textbf{836}, 50 (2017).
\newblock \doi{10.3847/1538-4357/836/1/50}

\bibitem{1974AnPhy..82..548W}
R.~{Wald}, Annals of Physics \textbf{82}, 548 (1974).
\newblock \doi{10.1016/0003-4916(74)90125-0}

\bibitem{2017JCAP...04..009A}
A.~{Aab}, P.~{Abreu}, M.~{Aglietta}, I.A. {Samarai}, I.F.M. {Albuquerque},
  I.~{Allekotte}, A.~{Almela}, J.~{Alvarez Castillo}, J.~{Alvarez-Mu{\~n}iz},
  G.A. {Anastasi}, et~al., JCAP \textbf{4}, 009 (2017).
\newblock \doi{10.1088/1475-7516/2017/04/009}

\bibitem{2016uhec.confa0016A}
R.~{Abbasi}, J.~{Bellido}, J.~{Belz}, V.~{de Souza}, W.~{Hanlon}, D.~{Ikeda},
  J.P. {Lundquist}, P.~{Sokolskypt}, T.~{Stroman}, Y.~{Tameda}, Y.~{Tsunesada},
  M.~{Unger}, A.~{Yushkov}, in \emph{Proceedings of International Symposium for
  Ultra-High Energy Cosmic Rays (UHECR2014), id.010016, pp.} (2016), p. 010016.
\newblock \doi{10.7566/JPSCP.9.010016}

\bibitem{2011APh....34..620A}
R.~{Aloisio}, V.~{Berezinsky}, A.~{Gazizov}, Astroparticle Physics \textbf{34},
  620 (2011).
\newblock \doi{10.1016/j.astropartphys.2010.12.008}

\bibitem{2003ApJ...591.1220L}
K.~{Lodders}, Astrophys. J. \textbf{591}, 1220 (2003).
\newblock \doi{10.1086/375492}

\bibitem{2006PhRvC..74d4902W}
K.~{Werner}, F.M. {Liu}, T.~{Pierog}, Phys. Rev. C \textbf{74}(4), 044902
  (2006).
\newblock \doi{10.1103/PhysRevC.74.044902}

\bibitem{2007APh....27...61A}
D.~{Allard}, E.~{Parizot}, A.V. {Olinto}, Astroparticle Physics \textbf{27}, 61
  (2007).
\newblock \doi{10.1016/j.astropartphys.2006.09.006}

\bibitem{1976ApJ...205..638P}
J.L. {Puget}, F.W. {Stecker}, J.H. {Bredekamp}, Astrophys. J. \textbf{205}, 638
  (1976).
\newblock \doi{10.1086/154321}

\bibitem{2015PhRvD..92f3014H}
D.~{Harari}, S.~{Mollerach}, E.~{Roulet}, Phys. Rev. D \textbf{92}(6), 063014
  (2015).
\newblock \doi{10.1103/PhysRevD.92.063014}

\bibitem{1966PhRvL..16..748G}
K.~{Greisen}, Physical Review Letters \textbf{16}, 748 (1966).
\newblock \doi{10.1103/PhysRevLett.16.748}

\bibitem{1966JETPL...4...78Z}
G.T. {Zatsepin}, V.A. {Kuz'min}, Soviet Journal of Experimental and Theoretical
  Physics Letters \textbf{4}, 78 (1966)

\bibitem{2018PrPNP..98...85M}
S.~{Mollerach}, E.~{Roulet}, Progress in Particle and Nuclear Physics
  \textbf{98}, 85 (2018).
\newblock \doi{10.1016/j.ppnp.2017.10.002}

\bibitem{1985PhRvD..31..564H}
C.T. {Hill}, D.N. {Schramm}, Phys. Rev. D \textbf{31}, 564 (1985).
\newblock \doi{10.1103/PhysRevD.31.564}

\bibitem{1988A&A...199....1B}
V.S. {Berezinskii}, S.I. {Grigor'eva}, Astron. Astrophys.p \textbf{199}, 1
  (1988)

\bibitem{2016NPPP..279..145S}
H.~{Sagawa}, {Telescope Array Collaboration}, Nuclear and Particle Physics
  Proceedings \textbf{279}, 145 (2016).
\newblock \doi{10.1016/j.nuclphysbps.2016.10.021}

\bibitem{2001PhRvL..86.2942L}
J.M. {Longuski}, E.~{Fischbach}, D.J. {Scheeres}, Physical Review Letters
  \textbf{86}, 2942 (2001).
\newblock \doi{10.1103/PhysRevLett.86.2942}

\bibitem{Whittaker}
E.T. Whittaker, G.N. Watson, \emph{A course of Modern Analysis} (Cambridge
  University Press, Cambridge, 1927)

\bibitem{Hagihara}
Y.~Hagihara, Japanese Journal of Astronomy and Geophysics \textbf{8}, 67 (1931)

\bibitem{2009CQGra..26k4002L}
G.~{Lovelace}, Classical and Quantum Gravity \textbf{26}(11), 114002 (2009).
\newblock \doi{10.1088/0264-9381/26/11/114002}

\bibitem{2016CQGra..33q5001S}
J.O. {Shipley}, S.R. {Dolan}, Classical and Quantum Gravity \textbf{33}(17),
  175001 (2016).
\newblock \doi{10.1088/0264-9381/33/17/175001}

\bibitem{2016PhRvD..94d4038D}
S.R. {Dolan}, J.O. {Shipley}, Phys. Rev. D \textbf{94}(4), 044038 (2016).
\newblock \doi{10.1103/PhysRevD.94.044038}

\bibitem{2006MNRAS.368.1515E}
P.~{Erdo{\v g}du}, J.P. {Huchra}, O.~{Lahav}, M.~{Colless}, R.M. {Cutri},
  E.~{Falco}, T.~{George}, T.~{Jarrett}, D.H. {Jones}, C.S. {Kochanek},
  L.~{Macri}, J.~{Mader}, N.~{Martimbeau}, M.~{Pahre}, Q.~{Parker},
  A.~{Rassat}, W.~{Saunders}, Mon. Not. R. Astron. Soc. \textbf{368}, 1515
  (2006).
\newblock \doi{10.1111/j.1365-2966.2006.10243.x}

\bibitem{2011ARA&A..49..119K}
K.~{Kotera}, A.V. {Olinto}, Ann. Rev. Astron. Astrophys. \textbf{49}, 119
  (2011).
\newblock \doi{10.1146/annurev-astro-081710-102620}

\bibitem{2014ApJ...781..119P}
R.~{Perna}, P.~{Duffell}, M.~{Cantiello}, A.I. {MacFadyen}, Astrophys. J.
  \textbf{781}, 119 (2014).
\newblock \doi{10.1088/0004-637X/781/2/119}

\bibitem{2011MNRAS.415.3824S}
N.~{Seto}, T.~{Muto}, Mon. Not. R. Astron. Soc. \textbf{415}, 3824 (2011).
\newblock \doi{10.1111/j.1365-2966.2011.18988.x}

\bibitem{2018arXiv180605820M}
I.~{Mandel}, A.~{Farmer}, ArXiv e-prints  (2018)

\bibitem{DiskDeadZone}
R.~{Martin}, C.~{Nixon}, F.G. {Xie}, A.~{King}, ArXiv e-prints  (2018)

\bibitem{2017MNRAS.465.4406K}
S.S. {Kimura}, S.Z. {Takahashi}, K.~{Toma}, Mon. Not. R. Astron. Soc.
  \textbf{465}, 4406 (2017).
\newblock \doi{10.1093/mnras/stw3036}

\bibitem{2017ApJ...839L...7D}
S.E. {de Mink}, A.~{King}, Astrophys. J. Lett. \textbf{839}, L7 (2017).
\newblock \doi{10.3847/2041-8213/aa67f3}

\bibitem{2017ApJ...848L..12A}
B.P. {Abbott}, R.~{Abbott}, T.D. {Abbott}, F.~{Acernese}, K.~{Ackley},
  C.~{Adams}, T.~{Adams}, P.~{Addesso}, R.X. {Adhikari}, V.B. {Adya}, et~al.,
  Astrophys. J. Lett. \textbf{848}, L12 (2017).
\newblock \doi{10.3847/2041-8213/aa91c9}

\bibitem{2017ApJ...850L..35A}
A.~{Albert}, M.~{Andr{\'e}}, M.~{Anghinolfi}, M.~{Ardid}, J.J. {Aubert},
  J.~{Aublin}, T.~{Avgitas}, B.~{Baret}, J.~{Barrios-Mart{\'{\i}}}, S.~{Basa},
  et~al., Astrophys. J. Lett. \textbf{850}, L35 (2017).
\newblock \doi{10.3847/2041-8213/aa9aed}

\bibitem{2005A&A...434..133W}
J.V. {Wall}, C.A. {Jackson}, P.A. {Shaver}, I.M. {Hook}, K.I. {Kellermann},
  Astron. Astrophys.p \textbf{434}, 133 (2005).
\newblock \doi{10.1051/0004-6361:20041786}

\bibitem{Abreu:2013kif}
P.~Abreu, et~al., JCAP \textbf{1305}(05), 009 (2013).
\newblock \doi{10.1088/1475-7516/2013/05/009}

\bibitem{2015ApJ...804...15A}
A.~{Aab}, P.~{Abreu}, M.~{Aglietta}, E.J. {Ahn}, I.A. {Samarai}, I.F.M.
  {Albuquerque}, I.~{Allekotte}, J.~{Allen}, P.~{Allison}, A.~{Almela}, et~al.,
  Astrophys. J. \textbf{804}, 15 (2015).
\newblock \doi{10.1088/0004-637X/804/1/15}

\bibitem{1984ARA&A..22..425H}
A.M. {Hillas}, Ann. Rev. Astron. Astrophys. \textbf{22}, 425 (1984).
\newblock \doi{10.1146/annurev.aa.22.090184.002233}

\bibitem{2017Sci...358.1299D}
Y.~{Dallilar}, S.S. {Eikenberry}, A.~{Garner}, R.D. {Stelter}, A.~{Gottlieb},
  P.~{Gandhi}, P.~{Casella}, V.S. {Dhillon}, T.R. {Marsh}, S.P. {Littlefair},
  L.~{Hardy}, R.~{Fender}, K.~{Mooley}, D.J. {Walton}, F.~{Fuerst},
  M.~{Bachetti}, A.J. {Castro-Tirado}, M.~{Charcos}, M.L. {Edwards}, N.M.
  {Lasso-Cabrera}, A.~{Marin-Franch}, S.N. {Raines}, K.~{Ackley}, J.G.
  {Bennett}, A.J. {Cenarro}, B.~{Chinn}, H.V. {Donoso}, R.~{Frommeyer},
  K.~{Hanna}, M.D. {Herlevich}, J.~{Julian}, P.~{Miller}, S.~{Mullin}, C.H.
  {Murphey}, C.~{Packham}, F.~{Varosi}, C.~{Vega}, C.~{Warner}, A.N.
  {Ramaprakash}, M.~{Burse}, S.~{Punnadi}, P.~{Chordia}, A.~{Gerarts}, H.~{de
  Paz Mart{\'{\i}}n}, M.M. {Calero}, R.~{Scarpa}, S.F. {Acosta}, W.M.
  {Hern{\'a}ndez S{\'a}nchez}, B.~{Siegel}, F.F. {P{\'e}rez}, H.D. {Viera
  Mart{\'{\i}}n}, J.A. {Rodr{\'{\i}}guez Losada}, A.~{Nu{\~n}ez},
  {\'A}.~{Tejero}, C.E. {Mart{\'{\i}}n Gonz{\'a}lez}, C.C. {Rodr{\'{\i}}guez},
  J.~{Molg{\'o}}, J.E. {Rodriguez}, J.I.F. {C{\'a}ceres}, L.A.
  {Rodr{\'{\i}}guez Garc{\'{\i}}a}, M.H. {Lopez}, R.~{Dominguez},
  T.~{Gaggstatter}, A.C. {Lavers}, S.~{Geier}, P.~{Pessev}, A.~{Sarajedini},
  Science \textbf{358}, 1299 (2017).
\newblock \doi{10.1126/science.aan0249}

\bibitem{2013PhRvD..88h4033Z}
F.~{Zhang}, B.~{Szil{\'a}gyi}, Phys. Rev. D \textbf{88}(8), 084033 (2013).
\newblock \doi{10.1103/PhysRevD.88.084033}

\bibitem{2015CQGra..32f5002B}
A.~{Bohn}, W.~{Throwe}, F.~{H{\'e}bert}, K.~{Henriksson}, D.~{Bunandar}, M.A.
  {Scheel}, N.W. {Taylor}, Classical and Quantum Gravity \textbf{32}(6), 065002
  (2015).
\newblock \doi{10.1088/0264-9381/32/6/065002}

\bibitem{2015PhRvD..91h3001B}
J.~{Brink}, M.~{Geyer}, T.~{Hinderer}, Phys. Rev. D \textbf{91}(8), 083001
  (2015).
\newblock \doi{10.1103/PhysRevD.91.083001}

\end{thebibliography}

\end{document}